\documentclass[fleqn,usenatbib]{mnras}
\usepackage{newtxtext,newtxmath}
\usepackage{longtable}
\usepackage[T1]{fontenc}

\DeclareRobustCommand{\VAN}[3]{#2}
\let\VANthebibliography\thebibliography
\def\thebibliography{\DeclareRobustCommand{\VAN}[3]{##3}\VANthebibliography}

\usepackage{graphicx}	% Including figure files
\usepackage{amsmath}	% Advanced maths commands
\title[The extremely young planetary nebula M\,3-27]{The extremely young planetary nebula M\,3-27: an analysis of its evolution, physical conditions and abundances}

\author[Francisco Ruiz-Escobedo, Miriam Pe\~na \& A.Valeria Beltr\'an-S\'anchez]{
Francisco Ruiz-Escobedo$^{1}$\thanks{E-mail: fdruiz@astro.unam.mx}
Miriam Pe\~na,$^{1}$ and Ana Valeria Beltr\'an-S\'anchez$^{1,2}$
\\
$^{1}$Instituto de Astronom\'ia, Universidad Nacional Aut\'onoma de M\'exico, Apdo. Postal 70264, Ciudad de M\'exico, M\'exico\\
$^{2}$Facultad de Ciencias, Universidad Nacional Aut\'onoma de M\'exico, Ciudad de M\'exico, M\'exico\\
}

\date{Accepted XXX. Received YYY; in original form ZZZ}

\pubyear{2022}

\begin{document}
\label{firstpage}
\pagerange{\pageref{firstpage}--\pageref{lastpage}}
\maketitle

\begin{abstract}
Spectrophotometric data of the young planetary nebula M\,3-27, from 2004 to 2021, are presented and discussed. We corroborate that the \ion{H}{i}  Balmer lines present features indicating they are emitted by the central star, therefore \ion{He}{i} lines were used to correct line fluxes by effects of reddening. Important variability on the nebular emission lines between 1964 to 2021, probably related to density changes in the nebula, is reported. Diagnostic diagrams to derive electron temperatures and densities have been constructed. The nebula shows a very large density contrast with an inner density of the order of 10$^{7}$  cm$^{-3}$  and an outer density of about $10^3 - 10^4$ cm$^{-3}$. With these values of density, electron temperatures of $16,000 - 18,000$ K have been found from collisionally excited lines. Due to the central star emits in the H$^+$ lines, ionic abundances relative to He$^+$  were calculated from collisionally excited and recombination lines, and scaled to H$^+$ by considering that He$^+$/H$^+$ $=$ He/H$ = 0.11$. ADF(O$^{+2}$) values were also determined. Total abundance values obtained indicate sub-solar abundances, similarly to what is found in other comparable objects like IC\,4997.
\end{abstract}

\begin{keywords}
planetary nebulae: individual: M\,3-27 -- ISM: abundances -- ISM: kinematics and dynamics
\end{keywords}

%%%%%%%%%%%%%%%%%%%%%%%%%%%%%%%%%%%%%%%%%%%%%%%%%%

%%%%%%%%%%%%%%%%% BODY OF PAPER %%%%%%%%%%%%%%%%%%

\section{Introduction}

Planetary nebulae (PNe) are formed by the ejection of parts of the stellar atmosphere during advanced stages of evolution of low-intermediate mass stars.
The ejecta, initially neutral or molecular, expands at velocities of about 10$-$20 km s$^{-1}$ and later it is photoionized by the energetic photons of the star when it evolves from the Asymptotic Giant Branch (AGB) phase  towards higher effective temperatures and reaches temperatures larger than 25,000 K.
The stellar transition between the post-AGB and the PN phase is not well known, it is not smooth and substantial photometric and spectral changes might occur in few years. 

There are few stars known to be in this transition period, presenting very young planetary nebulae. Some of them are IC\,4997, the Stingray nebula, VV\,8, PM\,1-322, and a few  others.

In this work, we analyse spectrophotometric data of the very young and compact PN M\,3-27 whose interesting behaviour has been studied since 50 years ago \citep{kohoutek:1968, adams:1975, ahern:1978, barker:1978, feibelman:1985, miranda:1997, wesson:2005}.  The main characteristics of M\,3-27 are listed in Table \ref{tab:properties}. Previous works showed the presence of two density zones inside the nebula, an inner zone of $n_e \geq$ 10$^6$ cm$^{-3}$ and a compact outer zone of $n_e \sim$ 10$^3 - $10$^4$ cm$^{-3}$.

Given the high density of the nebula, it has been difficult to determine a proper value of $T_e$ using plasma line diagnostics. 
\citet{ahern:1978} estimated a value of $T_e = 16,400$ K by comparing the line ratios of [\ion{O}{iii}]$\lambda$5007 and [\ion{Ne}{iii}]$\lambda$3869 according to the methodology presented by \citet{ahern:1975}. \citet{wesson:2005} determined a $T_e = 13,000$ K using the line ratio [\ion{O}{iii}] $\lambda \lambda$4959/4363 by assuming a density $n_e = 10^7$ cm$^{-3}$, estimated from high-order \ion{H}{i} Balmer lines.

By means of position-velocity  diagrams (PVD), \citet{miranda:1997} studied the line profiles and kinematics of this object. They found that the H$\alpha$ line has a kinematic profile different than those of [\ion{S}{ii}] and  [\ion{N}{ii}] lines. H$\alpha$ shows a double peak, much more intense in the red than in the blue, and wide wings extending up to $\pm$1500 km s$^{-1}$, possible due to Rayleigh-Raman dispersion\footnote{Rayleigh and Raman scattering are processes of photon scattering by atoms or molecules. \citet{lee:2000} and \citet{arrieta:2003} suggested that the main mechanism of formation of very broad wings on H$\alpha$ lines in young planetary nebulae is the Raman scattering of near-Ly$\beta$ photons produced by a dense atomic \ion{H}{i} region near to the central star. However, photons may suffer a series of Raman and Rayleigh scattering (in which an incident H$\alpha$ photon is scattered as another H$\alpha$ photon) before leaving this dense neutral region.} occurring in a very dense zone near the central star. In addition, faint H$\alpha$ emission can be traced up to about $\pm$12 arcsec from the stellar position, which is not shown by the heavy element lines. These latter lines and radio continuum emission at 3.6 cm are originated in a compact inner shell not spatially resolved (FWHM $\leq$1.4 arcsec). \citet{miranda:1997} estimated a kinematic age lower than 580 yr for the inner shell of this young PN.

\begin{table*}
    \centering
    \caption{Characteristics of M\,3-27}
    \label{tab:properties}
    \begin{tabular}{llrccccc}
    \hline
    Names & Gal. coord. & RA (2000) & Dec (2000) & Size & Parallax (mas)  & $v_{rad}$ (LSR) & Comments \\
        &    (deg)    & (hr:min:s)  &  (deg:min:sec) &      & Gaia DR3 &  (km s$^{-1}$)     &         \\ \hline \hline       
    PN G043.6+11.6  & 43.3554 +11.698 & 18:27:48 & +14:29:06 & Stellar & 0.1468$^a$ &  $-$24$^b$ &  Wide H$\alpha$\\
    IRAS 18255*14.27 &  &   &   &   &   &   & Double peak \ion{H}{i} lines\\ \hline
\multicolumn{8}{l}{$^a$ \citet{gaia:2016,gaia:2023} $^b$ \citet{miranda:1997} }\\
\end{tabular}
\end{table*}
    
We have studied this object as part of a project devoted to analyse young and dense PNe, whose first results were published by \citet{ruiz-escobedo:2022}. The general purpose of the project is to determine physical conditions and chemical composition of these young nebulae, with special emphasis in determining the Abundance Discrepancy Factors (ADF) between measurements  from collisionally excited lines (CELs) and recombination lines ORLs\footnote{The ADF is defined as the ratio of the abundance derived from ORLs to the abundance derived from CELs: ADF(X$^{+i}$) $=$ X$^{+i}_{ORLs}$ / X$^{+i}_{CELs}$.}.

This paper is organized as follows: In \S 1 we present the introduction with the general characteristics of M\,3-27. In \S 2, observations and data reduction are described. In \S 3 we analyse the calibrated line intensities to determine reddening corrections based \ion{He}{i} lines and analyse the temporal variations of line intensities. In \S 4 the line profiles are analysed. In \S 5 physical conditions and chemical abundances from collisionally excited lines and from recombination lines are determined and discussed. Line profiles and  kinematics of the plasma are discussed in \S 6. In \S 7 we present the discussion of our results and our conclusions can be found in \S 8.

\section{Observations}

Spectrophotometric data of M\,3-27 were obtained at the Observatorio Astron\'omico Nacional San Pedro M\'artir (OAN-SPM) at Baja California, M\'exico, between the years 2004 to 2022. Two spectrographs of different resolution, attached to the 2.1-m telescope, were used: the Boller \& Chivens (B\&Ch) spectrograph  of low-medium dispersion (resolution R ($\lambda / \delta \lambda$) of 685 at 5000 \AA) and the REOSC-Echelle spectrograph of high dispersion (resolution R $\sim$ 16,000$-$18,000 at 5000 \AA). The log of observations is presented in Table \ref{tab:log_observations}.

\begin{table*}
    \centering
    \caption{Log of observations for M\,3-27}
    \label{tab:log_observations}
    \begin{tabular}{llccccclc}
    \hline
Obs. date & Spectrograph$^{(a)}$ & \#exp~$\times~t_{exp}$ & Slit size & \multicolumn{2}{c}{\underline{spectral resolution}}  & $\lambda$ range   & ~~~Binning   & Airmass\\
&   & (s)   &   & \AA/pix & R(5000\AA)    & \AA & pix$\times$pix~~~\AA $\times ''$  & (average) \\ \hline \hline
22/04/2004 & B\&Ch (300 l mm$^{-1})^{(b)}$ & 2 $\times$ 60 & 5'$\times$4" & 2.23 &567 & 3450$-$7400 & 2$\times$2~~~ 7.0$\times$0.35 & 1.04 \\
25/04/2004  & REOSC-Echelle & 2 $\times$ 900& 13.3''$\times$4"  & 0.34& 16,000 &3600$-$6800 & 2$\times$2~~~ 0.3$\times$0.35  & 1.13 \\
30/06/2019  &  REOSC-Echelle & 3 $\times$ 1800& 13.3''$\times$2" & 0.30& 18,000 & 3600$-$7200 & 1$\times$1~~~  0.2$\times$0.18 & 1.05 \\
04/05/2021& B\&Ch (600 l mm$^{-1}$)& 3 $\times$ 900 & 5'$\times$4"& 1.54& 685& 3800$-$5950    & 2$\times$2~~~  5.4$\times$0.35 & 1.07 \\
05/05/2021& B\&Ch (600 l mm$^{-1}$)& 2 $\times$ 900 & 5'$\times$4" & 1.54&685 & 5200$-$7300  & 2$\times$2~~~ 5.4$\times$0.35& 1.07 \\
06/05/2021& B\&Ch (300 l mm$^{-1}$)& 2 $\times$ 300& 5'$\times$4" & 2.23&567 & 3450$-$7400   & 2$\times$2~~~  7.0$\times$0.35 & 1.24 \\
02/10/2021 & REOSC-Echelle  & 3 $\times$ 1800& 13.3"$\times$4" & 0.34&16,000 & 3600$-$7200 & 2$\times$2~~~ 0.3$\times$0.35     & 1.25 \\
16/08/2022 & REOSC-Echelle & 3 $\times$ 1800 & 13.3"$\times$2" & 0.30&18,000 & 3600--7200 & 1$\times$1~~~ 0.2$\times$0.18& 1.28 \\
\hline
\multicolumn{8}{l}{(a) The P.A. of the slit for B\&Ch observations was always E-W, for all REOSC-Echelle  it was E-W, except for 2021 when it was N-S.}\\
\multicolumn{6}{l}{(b) In parenthesis, the grating used.}
\end{tabular}
\end{table*}

\subsection{Data reduction}

\textsc{iraf} \footnote{\textsc{iraf} is distributed by the National Optical Observatory, which is operated by the Associated Universities for Research in Astronomy, Inc., under contract to the National Science Foundation.} tasks were used for data reduction.
The 2D B\&Ch frames were bias subtracted and flat fielded. Spectra were extracted (extraction window width included all the observed flux) and wavelength and flux calibrated. A Cu-Ar lamp, observed after each exposure, was used for wavelength calibration of B\&Ch observations. The standard stars Feige\,34, Feige\,66 and BD+30\,2642 were observed each night for flux calibration with a slit width of 500 $\mu$m (6.6 arcsec), in order to include all the stellar flux.  

2D REOSC-Echelle spectra were bias subtracted but no flat fielding was applied. Spectra were extracted from all the available orders, the extraction window included all the flux. Wavelength calibration was applied  with a Th-Ar lamp obtained after each scientific observation. Standard stars from the list by \citet{hamuy:1992} were observed for flux calibration. 

All observations were also corrected by effects of atmospheric extinction, using the curve determined by \citet{schuster:2001} for the OAN-SPM sky.

Calibrated spectra of different epochs have been used to determine variations of the emission lines, to analyse the line profiles, to determine the reddening, the physical conditions and the nebular ionic and total abundances. Fluxes from available lines in each spectrum were measured using \textsc{iraf}'s \textit{splot} task and normalized to the flux of H$\beta$ (H$\beta = 100$). Errors in line fluxes were estimated following \citet{tresse:1999} procedure. For REOSC-Echelle spectra, FWHM (full width at half-maximum) were also measured and corrected by effects of instrumental and thermal broadening (assuming the electron  temperature derived in each observation), subtracted in quadrature from the measured FWHM. For physical conditions and abundances, errors were estimated using Monte Carlo simulations, a set of 400 random points was generated following a normal distribution centred on each observed line intensity.

\section{Reddening correction}

In very high-density ionized gas, when the nebula is optically thick to H Lyman series and partially thick to H Balmer series, particularly thick to H$\beta$, the photons will be self-absorbed and will reappear in a cascade as P$\alpha$, H$\alpha$ and Ly$\alpha$  photons. Thus the H$\alpha$ photons emitted  will increase and H$\beta$ photons will decrease, increasing the observed H$\alpha$/H$\beta$ intensity ratio. Then such a ratio is not useful to determine the reddening. \citet{capriotti:1964b} calculated the theoretical values of the Balmer series in case of self-absorption for different optical depths. We have used Capriotti's procedure to determine c(H$\beta$) for the different observations presented in the literature
\citep[][and ours in this paper]{kohoutek:1968, adams:1975, ahern:1978, barker:1978} finding that the c(H$\beta$) value seems to vary with time, increasing from 0.50$-$0.75 for the observations of the seventies to values larger than 1.35$-$1.95 in the years from 2004 to 2021, due to the changes in H$\alpha$/H$\beta$ value probably produced by changes in the emission of the central star.

Due to the above, we consider that such a procedure is not appropriate to determine the external reddening for the OAN-SPM observations. Instead we decided to determine it by using the \ion{He}{i} lines following the procedure recently proposed by \citet{zamora:2022}. 
Such a procedure consists in using the intensities of the optical \ion{He}{i} singlet lines $\lambda\lambda$4922, 6678, 7281 and the triplet $\lambda$5876 line, relative to $\lambda$6678, because their theoretical intensities  are unaffected by optical depth and density effects, even for densities as large as 10$^8$ cm$^{-3}$, contrary to what occurs with other \ion{He}{i} triplet lines \citep{rodriguez:2020, zamora:2022}.

The expression used to calculate c(H$\beta$) is:

%The c(H$\beta$) coefficient is determined from the general expression for the calculation of interstellar extinction:

\begin{equation}
    log (F_\lambda / F_{ref}) - log  (I_\lambda / I_{ref}) = -c(H\beta) (f(\lambda) - f(\lambda_{ref})).
\end{equation}
  %log (F$_\lambda$ / F$_{ref}$) $-$ log  (I$_\lambda$ / I$_{ref}$) = $-$c(H$\beta$) (f($\lambda$) $-$ f($\lambda_{ref}$)), 

\noindent where $F_\lambda$ are the observed fluxes, $I_\lambda$ are the de-reddened fluxes, and $f(\lambda)$ is the extinction law given by \citet{cardelli:1989}, assuming a ratio of total to selective extinction $R_V = 3.1$. The reference wavelength ($\lambda_{ref}$) is $\lambda$6678 \AA, which corresponds to the most intense of the \ion{He}{i} singlet lines. This line will be used to normalize the line intensities.

By using \textsc{PyNeb} \citet{luridiana:2015}, the theoretical intensities $I_\lambda$ were calculated for a density of 10$^7$ cm$^{-3}$ and a temperature of 10$^4$ K, the atomic parameters used are those by \citet{porter:2012, porter:2013}. 

\begin{figure*}
	\includegraphics[width=\columnwidth]{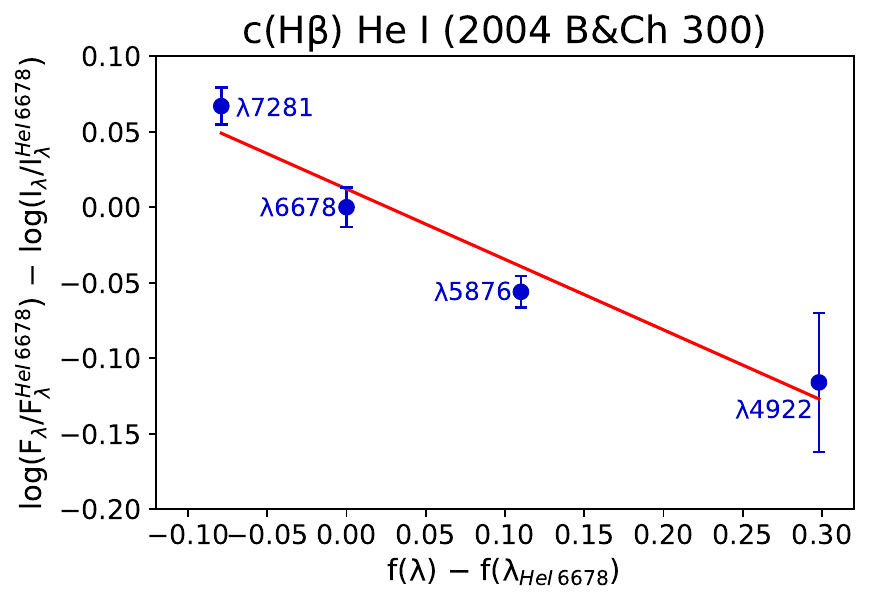}
	\includegraphics[width=\columnwidth]{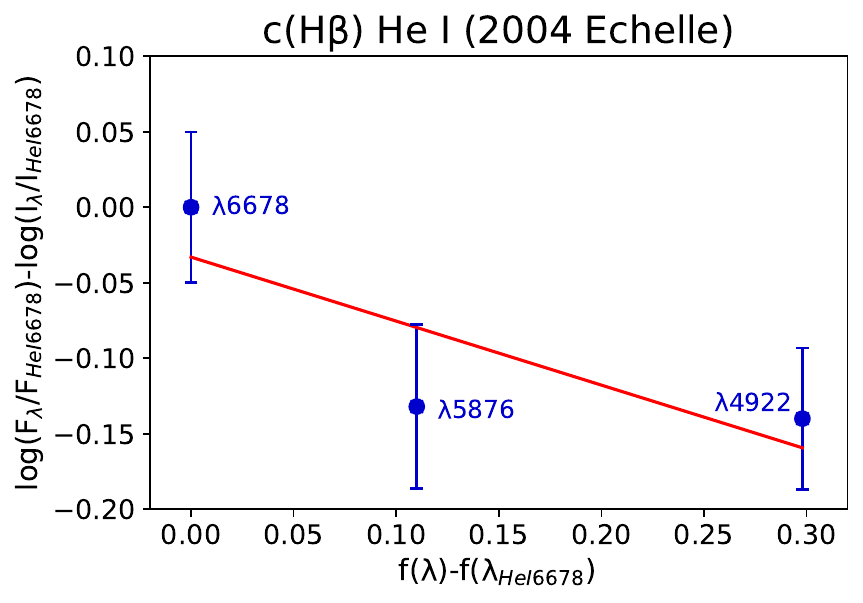}
	\includegraphics[width=\columnwidth]{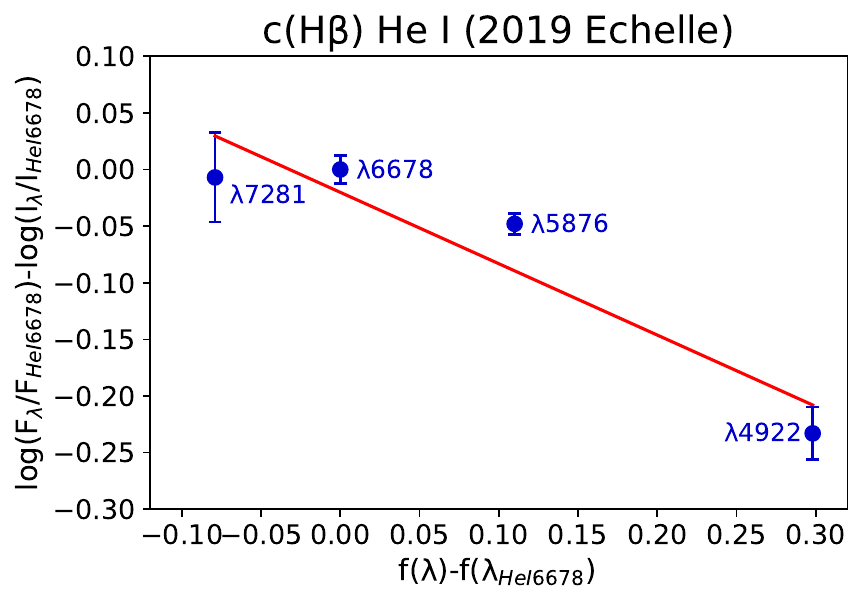}
	\includegraphics[width=\columnwidth]{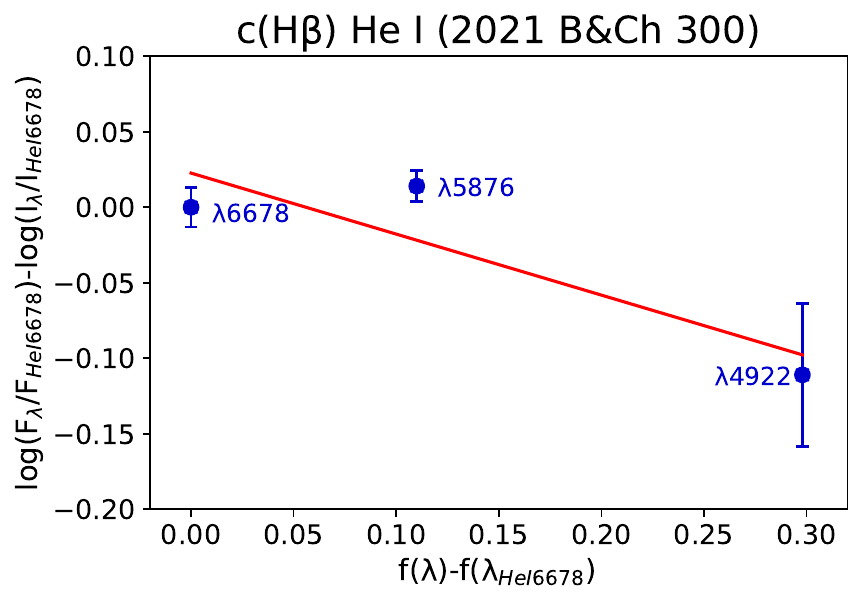}
        \includegraphics[width=\columnwidth]{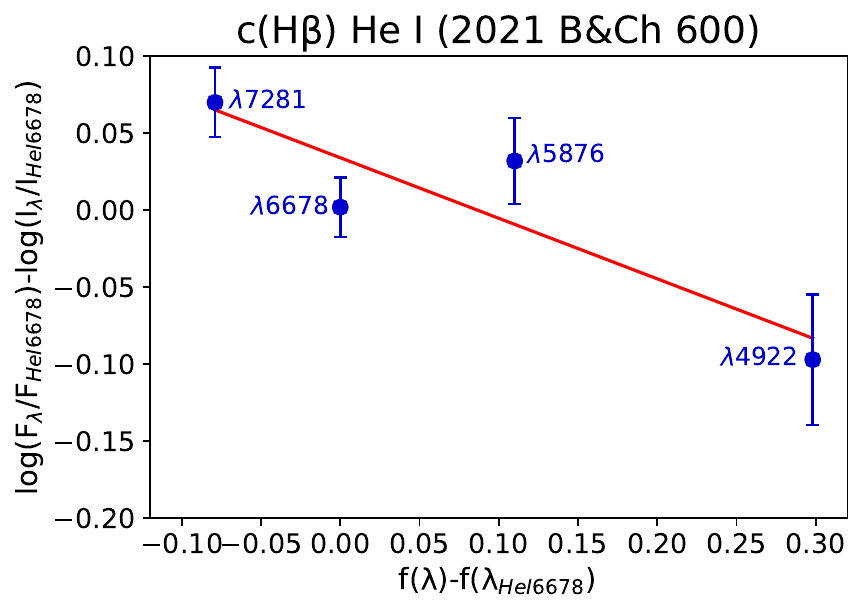}
	\includegraphics[width=\columnwidth]{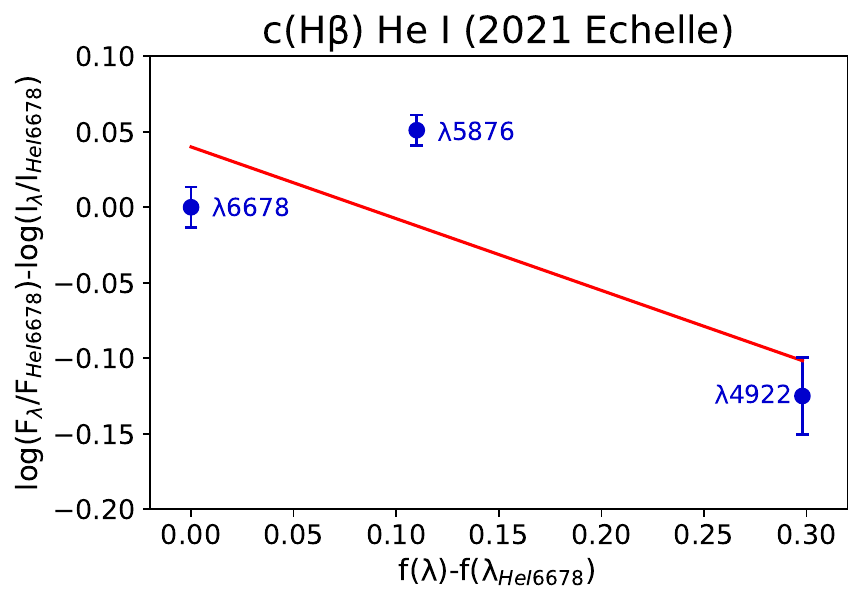}
    \caption{Extinction determination using the \ion{He}{i} lines. c(H$\beta$) is the slope of the calculated lines.} 
    \label{fig:hei_extinction}
\end{figure*}

 The ratios of \ion{He}{i} observed fluxes and dereddened fluxes for all the observations from 2004 to 2021, normalized to the line $\lambda$6678 \AA~ as said above, and their values of $f(\lambda)$ were plotted in the plane $log (F_\lambda / F_{ref}) - log  (I_\lambda / I_{ref})$ vs. $f(\lambda) - f(\lambda_{ref})$  (see Fig. \ref{fig:hei_extinction}).  A linear regression was applied to all the plots, deriving in each case the c(H$\beta$) value which is the slope of the calculated lines.

%These intensities and the expression given above, allowed us to calculate the c(H$\beta$) values for all the observations from 2004 to 2021. by performing a linear regression. A density of 10$^7$ cm$^{-3}$, adequate for the inner zone of M\,3-27, and the reddening law by \citet{cardelli:1989} was adopted. 

The results are presented in Table \ref{tab:c(Hbeta)} where the RMS of the regression is adopted as the error for the determined c(H$\beta$). The value  R$^2$ of the regression is given. For each observation, the c(H$\beta$) values derived this way were used to deredden the spectra.

\begin{table*}
    \centering
    \caption{Logarithmic reddening c(H$\beta$) from \ion{He}{I} lines for M3-27}
    \label{tab:c(Hbeta)}
    \begin{tabular}{llcccccc}
    \hline
    Year & Obs & c(H$\beta$) & RMS & R$^2$ & \ion{He}{I} observed lines\\ 
    \hline
    2004 & B\&Ch (300 l/mm)& 0.47 & 0.02 & 0.95 & 4922, 5876, 6678, 7281\\
    2004 & REOSC-Echelle & 0.42 & 0.04 & 0.66 & 4922, 5876, 6678 \\
    2019 & REOSC-Echelle & 0.63 & 0.03 & 0.89 & 4922, 5876, 6678, 7281 \\
    2021 & B\&Ch (300 l/mm) & 0.40 & 0.03 & 0.79 & 4922, 5876, 6678\\
    2021 & B\&Ch (600 l/mm) & 0.46 & 0.04 & 0.72 & 4922, 5876, 6678, 7281 \\
    2021 & REOSC-Echelle & 0.48 & 0.05 & 0.63 & 4922, 5876, 6678 \\
    \hline
%\multicolumn{6}{l}{The P.A. (B\&Ch) was E-W always. Echelle no se!}\\
\end{tabular}
\end{table*}

The observed and dereddened  fluxes relative to H$\beta$ and the FWHM of all the detected lines, for all the observations, are presented in a table on-line. In Table \ref{tab:line-intensities} we present an example of such a table.

\begin{table*}

\caption{Example: observed ($F_{\lambda}$) and dereddened ($I_{\lambda}$) fluxes normalized to H$\beta$ = 100,
their absolute errors, FWHM of line widths and expansion and heliocentric velocities$^a$.}

\begin{tabular}{lccccccccc}

\hline

Ion & $\lambda_0$ & $\lambda_{obs}$ & F$_\lambda$/F(H$\beta$) & $\delta($F$_\lambda$) & I$_\lambda$/I(H$\beta$) & $\delta($I$_\lambda$) & FWHM & $v_{exp}$ & $v_{rad}$ \\
  & (\AA) & (\AA) &    &  &   & & (\AA) & (km s$^{-1}$)  & (km s$^{-1}$) \\ \hline \hline

H25 & 3669.47 & 3668.66 & 0.28 & 0.11 & 0.46 & 0.19 & 0.23 & 9.59 & $-$66.35 \\
H24 & 3671.48 & 3670.67 & 0.51 & 0.17 & 0.83 & 0.27 & 0.26 & 10.79 & $-$66.31 \\
H23 & 3673.76 & 3672.90 & 0.61 & 0.27 & 0.97 & 0.43 & 0.52 & 21.31 & $-$70.27 \\
H22 & 3676.37 & 3675.50 & 0.73 & 0.21 & 1.17 & 0.34 & 0.71 & 28.99 & $-$70.71 \\
H21 & 3679.36 & 3678.50 & 0.96 & 0.22 & 1.53 & 0.36 & 0.22 & 9.15 & $-$70.41 \\
H20 & 3682.81 & 3681.92 & 1.13 & 0.27 & 1.80 & 0.43 & 0.41 & 16.87 & $-$72.13 \\
H19 & 3686.83 & 3686.03 & 1.31 & 0.39 & 2.10 & 0.62 & 0.33 & 13.38 & $-$65.22 \\
H18 & 3691.56 & 3690.68 & 1.06 & 0.36 & 1.68 & 0.58 & 0.25 & 10.25 & $-$71.39 \\
H17 & 3697.15 & 3696.39 & 1.81 & 0.51 & 2.88 & 0.81 & 0.49 & 20.08 & $-$61.39 \\
H16 & 3703.86 & 3703.03 & 2.12 & 0.48 & 3.40 & 0.77 & 0.59 & 23.79 & $-$66.94 \\
\ion{He}{i} & 3705.04 & 3704.10 & 0.92 & 0.35 & 1.47 & 0.57 & 0.52 & 20.88 & $-$75.75 \\
H15 & 3711.97 & 3711.14 & 2.21 & 0.31 & 3.54 & 0.50 & 0.76 & 30.71 & $-$66.72 \\
H14+[\ion{S}{iii}] & 3721.83 & 3720.94 & 4.01 & 0.34 & 6.41 & 0.56 & 0.69 & 27.79 & $-$72.10 \\
$[$\ion{O}{ii}] & 3726.03 & 3725.28 & 0.48 & 0.11 & 0.76 & 0.18 & 0.36 & 14.51 & $-$60.75 \\
$[$\ion{O}{ii}] & 3728.82 & 3728.08 & 0.29 & 0.11 & 0.46 & 0.17 & 0.84 & 33.88 & $-$59.58 \\
H13 & 3734.37 & 3733.52 & 3.12 & 0.26 & 4.95 & 0.43 & 0.73 & 29.21 & $-$67.92 \\
H12 & 3750.15 & 3749.34 & 3.84 & 0.55 & 6.07 & 0.88 & 0.80 & 31.95 & $-$64.84 \\
H11 & 3770.63 & 3769.81 & 4.37 & 0.30 & 6.80 & 0.49 & 0.42 & 16.52 & $-$64.96 \\ \hline

\multicolumn{10}{l}{$^a$ $F_{\lambda}$: observed flux, $\delta (F_{\lambda})$: error, $I_{\lambda}$: dereddened flux, $\delta (I_{\lambda})$: error, $v_{exp}$: expansion velocity,} \\
\multicolumn{10}{l}{$v_{rad}$: heliocentric velocity. The complete table, including all the observations, is on-line-material.}

\end{tabular}

\label{tab:line-intensities}

\end{table*}

\subsection{Temporal variations of line intensities}

\begin{figure*}
	\includegraphics[width=\columnwidth]{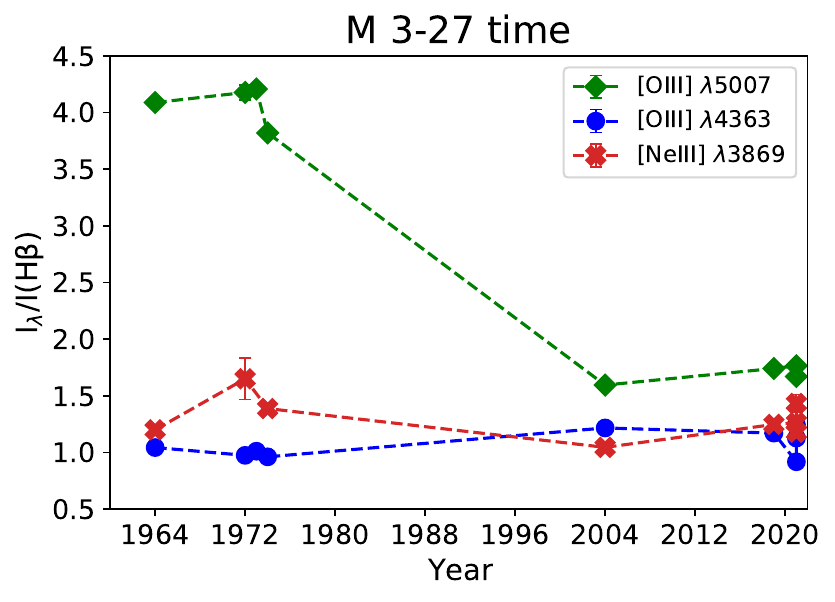}
	\includegraphics[width=\columnwidth]{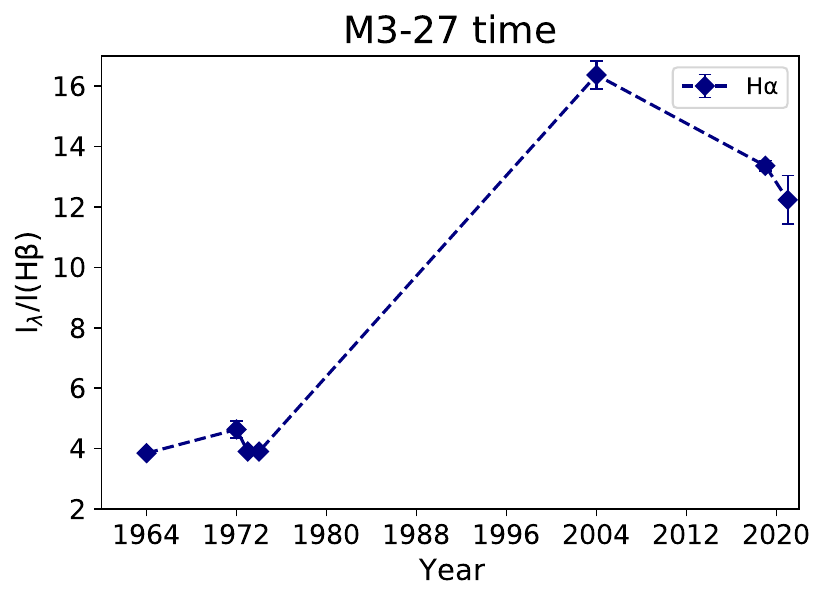}
	\includegraphics[width=\columnwidth]{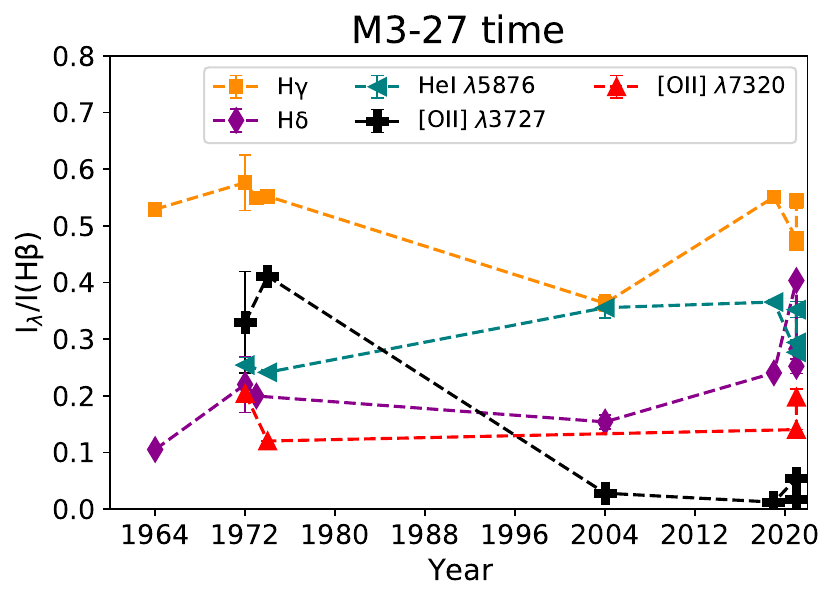}
    \includegraphics[width=\columnwidth]{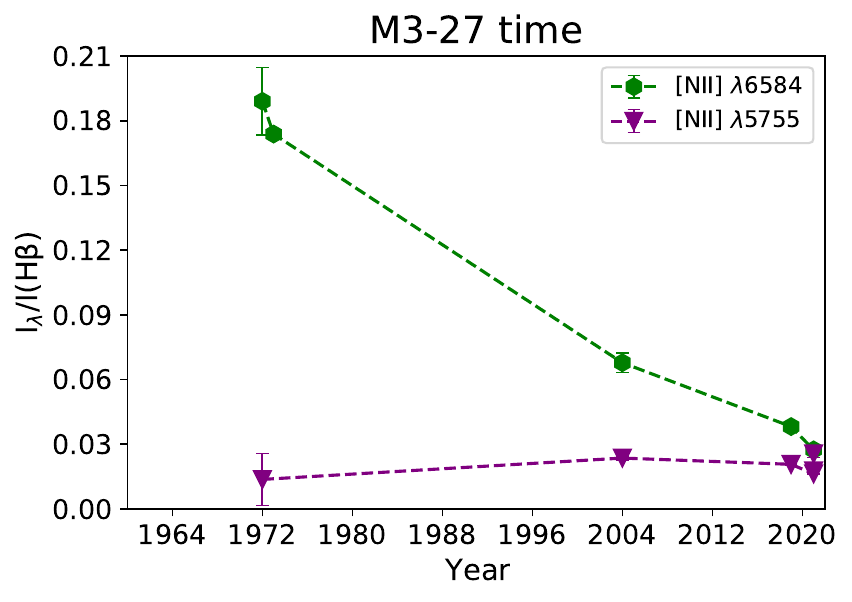}
    \caption{Temporal evolution of the dereddened intensities of important lines. In these plots, lines are normalized to H$\beta = 1$.}
    \label{fig:evolution}
\end{figure*}

Fig. \ref{fig:evolution} shows the temporal evolution of some important lines in the period from 1964 to 2021. The data from 1964 to 1978 were obtained from \citet{kohoutek:1968, adams:1975, ahern:1978} \citet{barker:1978}.
 In Fig. \ref{fig:evolution} (up, left) the evolution of lines [\ion{O}{iii}]$\lambda$5007, $\lambda$4363 and [\ion{Ne}{iii}]$\lambda$3869 relative to H$\beta$ are presented. It is observed that the nebular line [\ion{O}{iii}]$\lambda$5007  diminished by a factor of 2.7 in this period, while the auroral line [\ion{O}{iii}]$\lambda$4363  and the [\ion{Ne}{iii}]$\lambda$3869 line do not show significant variations. This is a consequence of the critical density ($n_{crit}$) that is lower for $\lambda$5007 than for the other lines (see Table \ref{tab:critical-densities}).  In the case of \ion{H}{i} Balmer lines (Fig. \ref{fig:evolution} (up, right)) impressive changes are found, the H$\alpha$/H$\beta$ ratio increased by a factor of 4 in the period of 50 years, from $\sim 4$ during 1970 to $\sim 12$ in 2021, while H$\gamma$ and H$\delta$ show no significant variations relative to H$\beta$. This confirms that H$\alpha$ is partially emitted by the central star and its variations can be attributed to changes in the stellar emission.

The behaviour of other heavy element lines is shown in Fig. \ref{fig:evolution} (low, left) where the most large variation is presented by [\ion{O}{II}]$\lambda$3727+ nebular lines (the addition of [\ion{O}{II}]$\lambda$3726 and $\lambda$3729 lines) which decreased by a factor of 4 in the period from 1970 to 2020, from $\sim 0.4 \rm{H}\beta$ in 1970 to $< 0.1 \rm{H}\beta$ in 2020, while the [\ion{O}{II}] $\lambda$7325+ auroral lines (the addition of [\ion{O}{II}]$\lambda$7319, $\lambda$7320, $\lambda$7329,and $\lambda$7330 lines) show small variations. Other lines present smaller changes. \ion{He}{i} $\lambda$5876 increased by a factor of 1.4 in 2004  while [\ion{S}{ii}] $\lambda$6717  did not vary significantly.  In Fig. \ref{fig:evolution} (low, right) the behaviour of [\ion{N}{ii}] $\lambda$6584 nebular line and $\lambda$5755 auroral line are shown, in this case the nebular line $\lambda$6584 had varied significantly while the auroral line $\lambda$5755 presents small changes. 

 All the above is an effect  of the critical densities that are lower for the lines [\ion{O}{ii}]$\lambda$3727+ and [\ion{N}{ii}]$\lambda$6584 than for the others. The critical densities derived from \textsc{PyNeb} for the lines mentioned here are tabulated in Table \ref{tab:critical-densities}. The atomic parameters used are listed in  Table \ref{tab:atomic-parameters}, and the temperature assumed was 17,000 K which is an average value of the temperatures derived for the different observations (see Table \ref{tab:phys-cond}).

\begin{figure*}
	\includegraphics[width=\columnwidth]{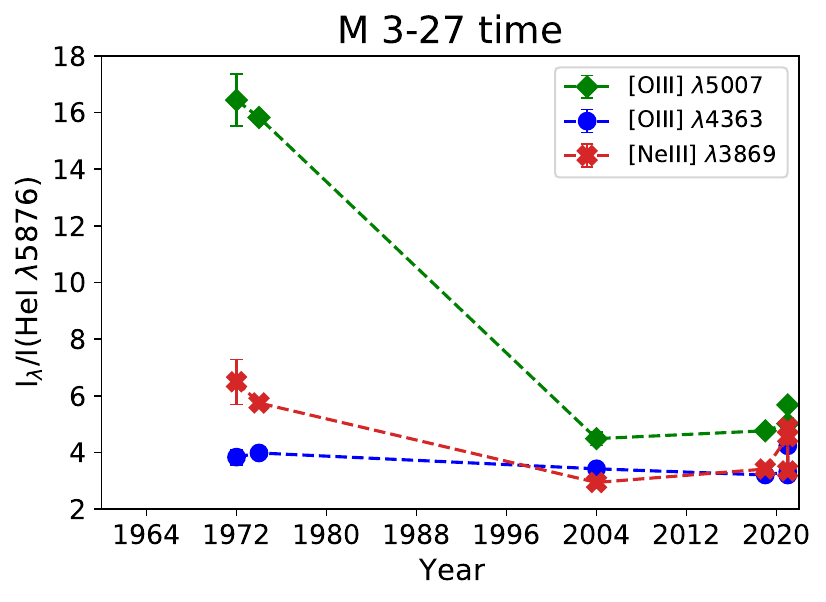}
	\includegraphics[width=\columnwidth]{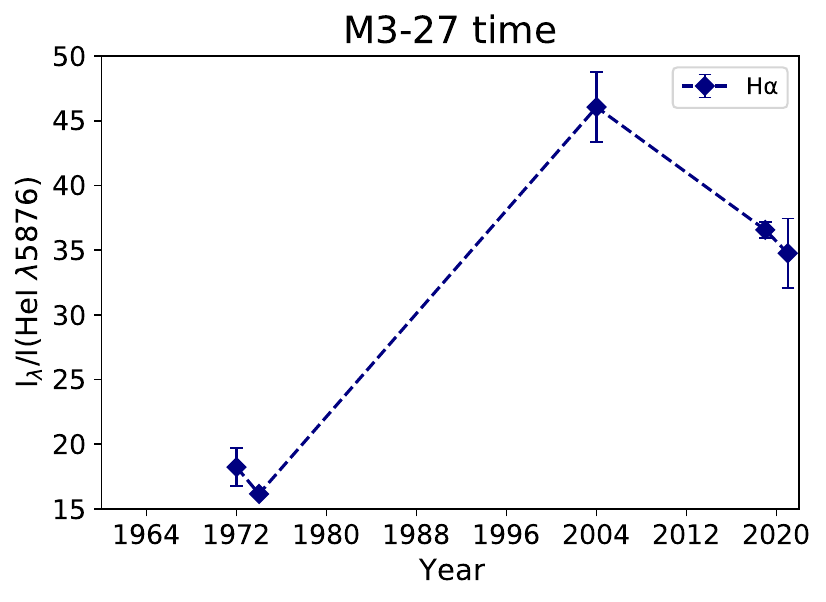}
	\includegraphics[width=\columnwidth]{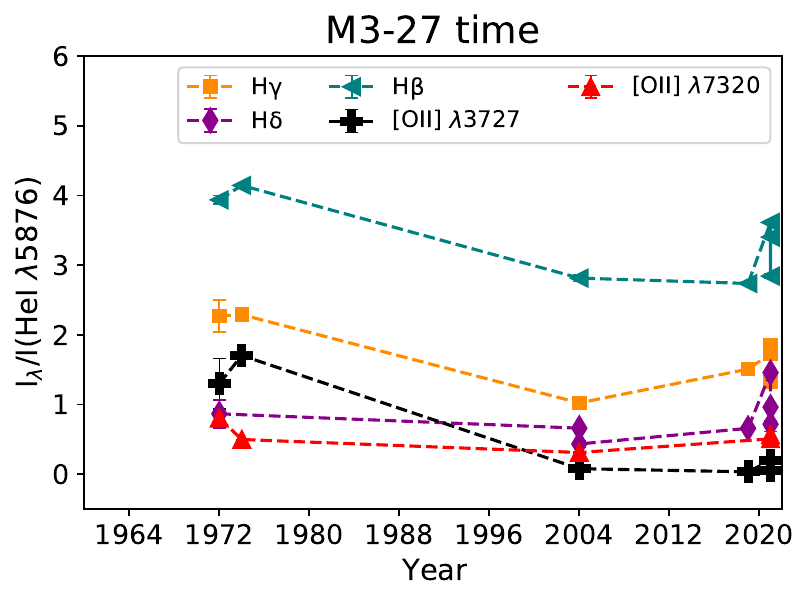}
    \includegraphics[width=\columnwidth]{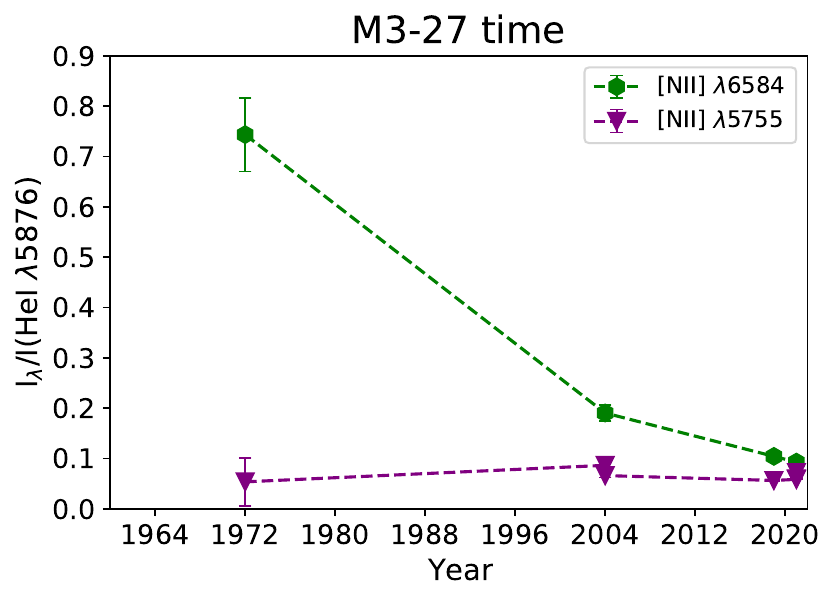}
    \caption{Temporal evolution of the dereddened intensities of important lines relatives to \ion{He}{i} line $\lambda$5876.}
    \label{fig:evolution_he}
\end{figure*}

As \ion{H}{i} lines show characteristics indicating they are emitted by the stellar atmosphere and present variations through time, we also analysed the evolution of line intensities relative to \ion{He}{i}  $\lambda$5876 line, which is emitted by the nebula.  In Figure \ref{fig:evolution_he}, we present the temporal evolution of the intensities of the same lines analysed above but normalized to the mentioned \ion{He}{i} line. For the observations from the 1970s, the \ion{He}{i} $\lambda$5876 line is only reported in the data from \citet{ahern:1978} and \citet{barker:1978}, however the line intensities show the same behaviour as it was found for the normalization to H$\beta$, highlighting the suppression of the emission of nebular lines of [\ion{O}{ii}], [\ion{N}{ii}] and [\ion{O}{iii}].
 
%\begin{figure}
 %   \centering
  %  \includegraphics[width=\columnwidth]{r1_r2_gm95.pdf}
   % \caption{R3 parameter behaviour in R1-R2 plane.}
 %%\end{figure}

\begin{table}
\centering
\caption{Critical electron densities ($n_{crit}$) for lines of interest at $T_e$= 1.7$\times$10$^4$ K.}
\label{tab:critical-densities}
\begin{tabular}{lcc} 
\hline
Ion & $\lambda$ & $n_{crit}$ $^a$ \\
    & (\AA) & (cm$^{-3}$) \\
\hline
%$[$\ion{O}{i}] & 6300 &  1.16$\times$10$^\text{6}$\\
$[$\ion{O}{ii}] & 3726 & 4.97$\times$10$^\text{3}$\\
$[$\ion{O}{ii}] & 3729 &  1.43$\times$10$^\text{3}$\\
$[$\ion{O}{ii}] & 7325 &  4.92$\times$10$^\text{6}$\\
$[$\ion{O}{iii}] & 4363 &  3.06$\times$10$^\text{7}$\\
$[$\ion{O}{iii}] & 5007 &  8.15$\times$10$^\text{5}$\\
$[$\ion{N}{ii}] & 5755 &  1.95$\times$10$^\text{7}$\\
$[$\ion{N}{ii}] & 6584 & 1.09$\times$10$^\text{5}$\\
$[$\ion{Ne}{iii}] & 3869 &  1.33$\times$10$^\text{7}$\\
$[$\ion{S}{ii}] & 6717 & 1.44$\times$10$^\text{3}$\\
$[$\ion{S}{ii}] & 6731 & 3.89$\times$10$^\text{3}$\\
$[$\ion{S}{ii}] & 4068 & 2.22$\times$10$^\text{6}$\\
$[$\ion{Ar}{iii}] & 5191 & 5.06$\times$10$^\text{7}$\\
$[$\ion{Ar}{iii}] & 7136 & 6.92$\times$10$^\text{6}$\\
$[$\ion{Ar}{iv}] & 4711 & 1.96$\times$10$^\text{4}$\\
$[$\ion{Ar}{iv}] & 4740 &  1.20$\times$10$^\text{5}$\\
$[$\ion{Ar}{iv}] & 7170 &  2.16$\times$10$^\text{7}$\\
$[$\ion{Cl}{iii}] & 5518 & 7.34$\times$10$^\text{3}$\\
$[$\ion{Cl}{iii}] & 5538 &  4.01$\times$10$^\text{4}$\\  \hline
\multicolumn{3}{l}{$^a$ Critical densities calculated with \textsc{PyNeb}}\\
\multicolumn{3}{l}{using atomic data presented in Appendix C.}
\end{tabular}
\end{table}

\section{The lines profiles}

\begin{figure}
    \centering
    \includegraphics[width=\columnwidth]{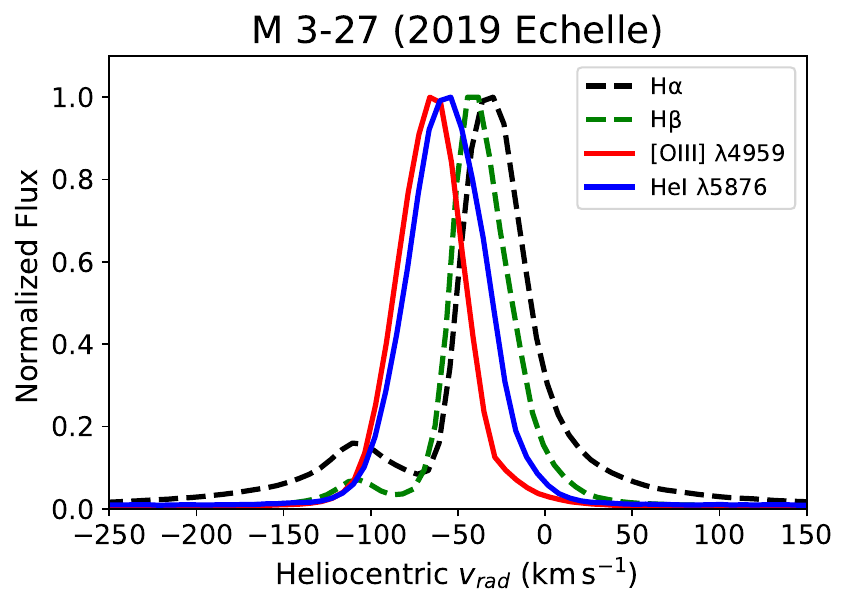}
    \caption{Profiles and heliocentric radial velocities of important lines.}
    \label{fig:profiles}
\end{figure}

In Fig. \ref{fig:profiles} the line profiles of H$\alpha$, H$\beta$, [\ion{O}{iii}]$\lambda$4959 and \ion{He}{i} $\lambda$5876 are shown from the 2019 high resolution REOSC-Echelle observations. 
It is found that both \ion{H}{i} lines show double peak with a faint component appearing to the blue.
The red component is very intense in comparison with the blue one. Both peaks are separated by about 70 km s$^{-1}$. Such a profile was already found by \citet{miranda:1997} who reported that H$\alpha$ exhibits a type III P-Cygni profile with two emissions at $v_{LSR}$  $-$84 and $-$7 km s$^{-1}$, separated by an absorption at $-$55 km s$^{-1}$. In addition H$\alpha$ shows very extended wings traced up to $\pm$1500 km s$^{-1}$. We have found this profile in all the Balmer lines at least  up to H$\delta$. The \ion{H}{i} line profiles are different from the profiles of \ion{He}{i} and heavy element lines which show a unique peak at a systemic velocity $v_{hel}$ of $\sim -70$ km s$^{-1}$ as shown in Fig. \ref{fig:profiles}. Such peak coincides better with the position of the absorption shown by the Balmer lines, which indicates that the Balmer lines  originate mainly in a different zone than the \ion{He}{i} and the heavy element lines, most probably in the atmosphere of the central star. This will complicate the calculus of abundances as it will be seen in the next sections.

\begin{figure}
    \centering
    \includegraphics[width=\columnwidth]{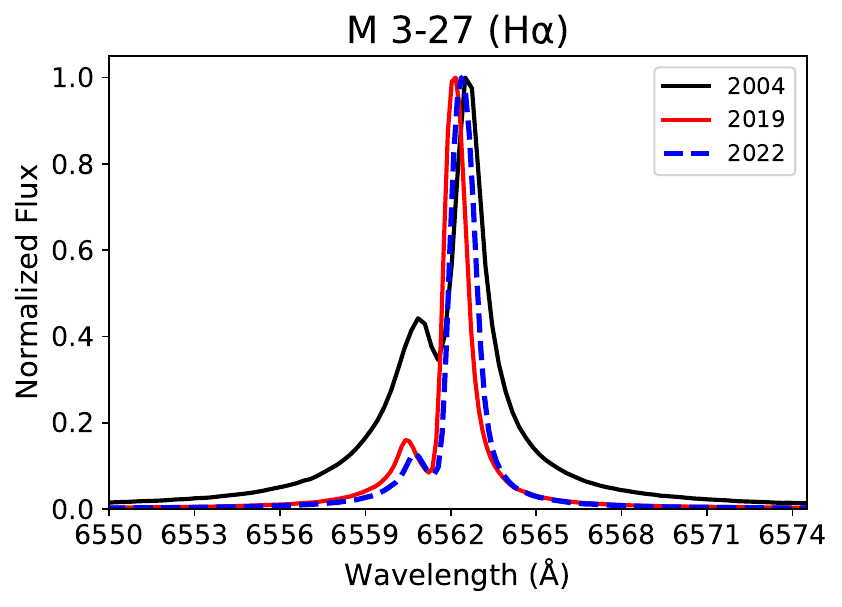}
    \caption{Variations of the stellar profile of H$\alpha$ with time. Notice that the blue component was more intense in 2004. H$\alpha$ profile from 2021 spectrum is not presented due it is saturated.}
    \label{fig:halpha-profiles}
\end{figure}

 An interesting fact found in this work is that the line profile of H$\alpha$ seems to have changed with time (see Fig. \ref{fig:halpha-profiles}). In particular, the blue component of the profile was more intense in 2004 than it was in the period 2019 to 2022. The observations reported by \citet{miranda:1997}, obtained in 1993, also shows a blue component similar to the ones of the 2019$-$2022 period, with an intensity of  about 0.2 the value for the red component. However \citet{tamura:1990} shows a single profile for H$\alpha$ from observations obtained  by K. M. Shibata previous to 1988.  

\section{Physical conditions and chemical abundances} \label{sec:physical_conditions}

\begin{figure*}
	\includegraphics[width=\columnwidth]{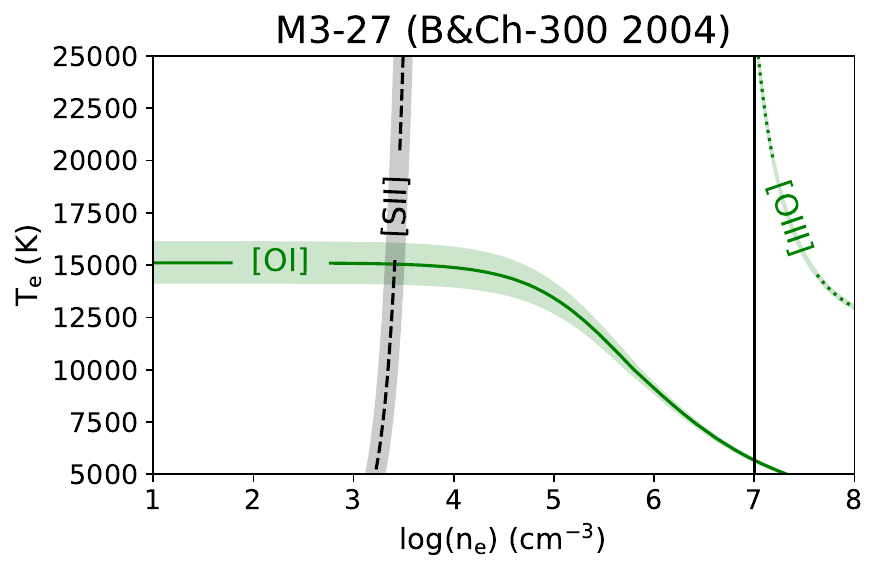}
	\includegraphics[width=\columnwidth]{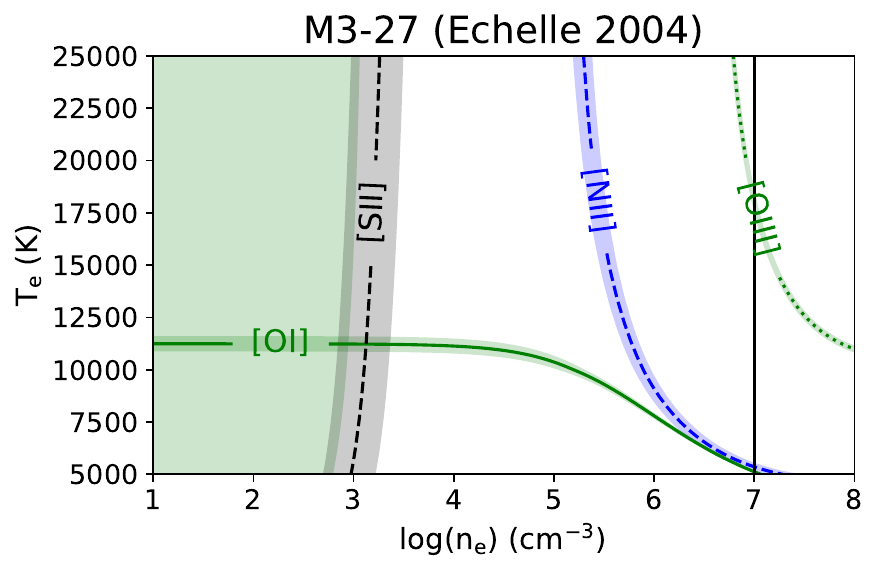}
	\includegraphics[width=\columnwidth]{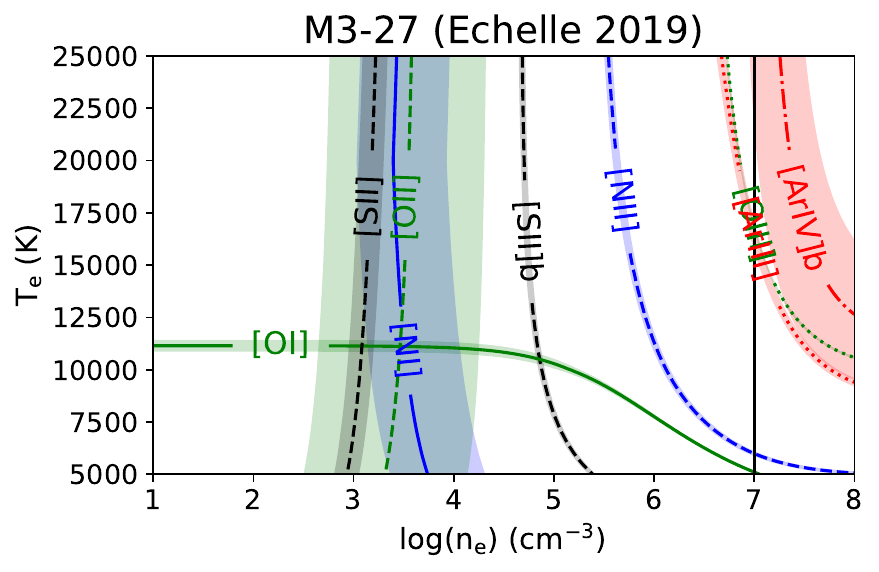}
    \includegraphics[width=\columnwidth]{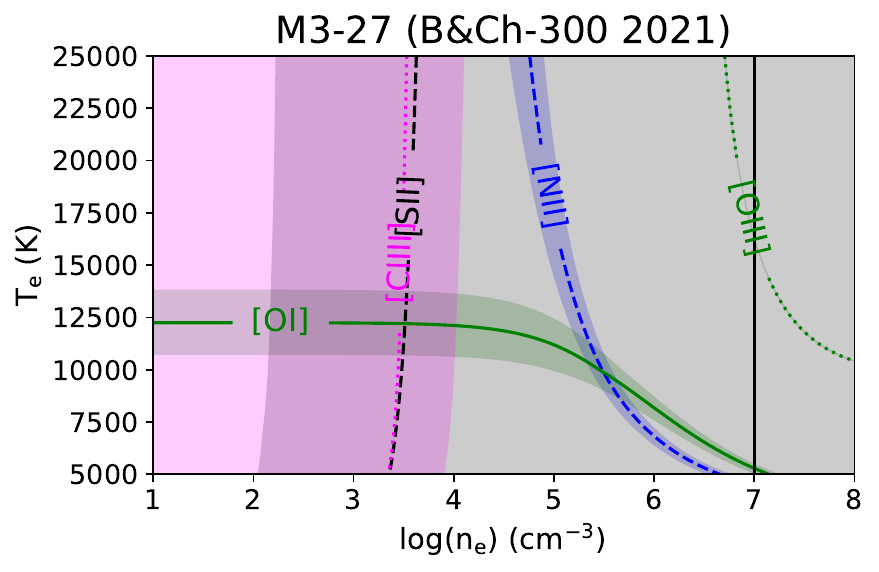}

    \includegraphics[width=\columnwidth]{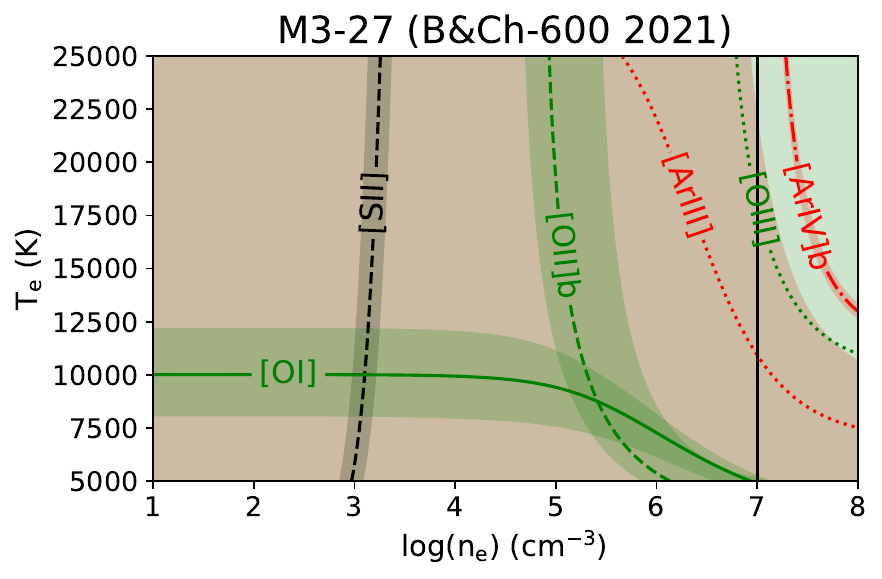}
    \includegraphics[width=\columnwidth]{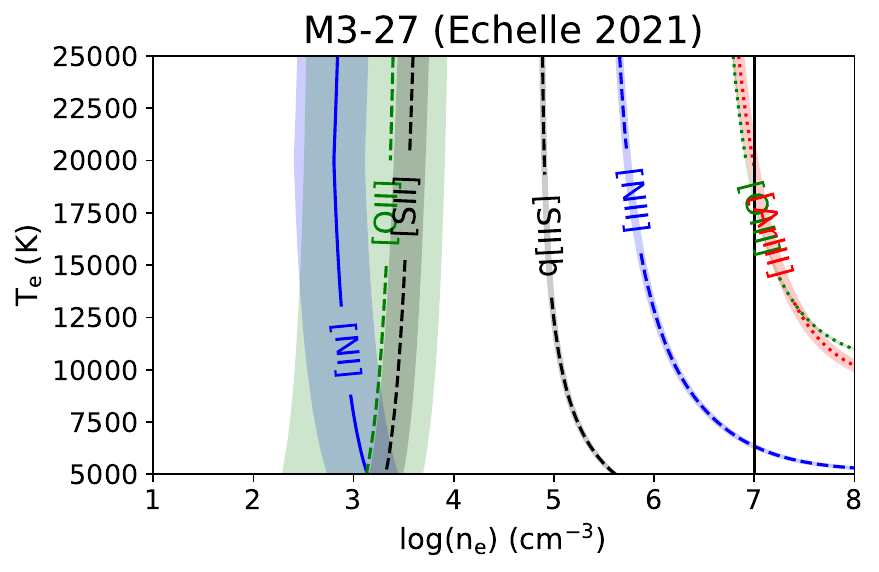}
    \caption{Diagnostic diagrams for density and temperature, derived with \textsc{PyNeb}. Sensitive CELs line ratios [\ion{O}{iii}]$\lambda\lambda$(5007+4959)/4363,  [\ion{N}{ii}]$\lambda\lambda$(6584+6548)/5755,  [\ion{Ar}{iii}] $\lambda\lambda$7136/5192, [\ion{S}{ii}]$\lambda \lambda$ 6716/6731, and [\ion{O}{ii}]$\lambda \lambda$3726/3729 are indicated. Coloured shadowed bands represent the 1 sigma rms error of each diagnostic. [\ion{O}{ii}]b, [\ion{S}{ii}]b and [\ion{Ar}{iv}]b represent the line ratios $\lambda \lambda$3727+/7325+, $\lambda \lambda$(6716+6731)/4068 and $\lambda \lambda$7170/4740, respectively. The vertical solid line in all diagrams represents a constant $n_e$ of $10^7$ cm$^{-3}$.}
    \label{fig:diags}
\end{figure*}

Physical conditions, $T_e$ and $n_e$, were calculated using \textsc{PyNeb} \citep{luridiana:2015}. This was done by adopting the line intensities relative to H$\beta$ presented in Table {\ref{tab:line-intensities}}, determined including all the observed emission (in the case of \ion{H}{i} Balmer lines both components were included), corrected by reddening and using the atomic data presented in Table \ref{tab:atomic-parameters}. Several plasma diagnostic intensity ratios from the collisionally excited lines (CELs) of different ions are available to determine electron temperature and densities. The behaviour of these diagnostic lines can be observed in the diagnostic diagrams (Figure \ref{fig:diags}), which were constructed using \textsc{PyNeb} for OAN-SPM observations between 2004 and 2021.

The diagnostic diagrams show that the ``traditional'' auroral-nebular line intensity ratios normally used to determine electron temperature, such as [\ion{O}{iii}]$\lambda\lambda$(5007+4959)/4363,  [\ion{N}{ii}]$\lambda\lambda$(6584+6548)/5755 and [\ion{Ar}{iii}] $\lambda\lambda$7136/5192, cannot be used with such a purpose in M\,3-27 due to such ratios are more indicative of the electron density. A density of the order of $n_e \sim 10^7$ cm$^{-3}$, higher than the critical density of some nebular lines, are derived from these ratios. On the other hand, the [\ion{S}{ii}] $\lambda \lambda$6731/6716 and [\ion{O}{ii}] $\lambda \lambda$3729/3726  intensity ratios indicate a much lower density ($n_e \sim 10^3$ cm$^{-3}$). It is clear from such diagrams that M\,3-27 has a large density contrast, being much denser in the inner zone.

The contribution of recombination to the intensity of the nebular line [\ion{N}{ii}] $\lambda$5755 was determined using the relation proposed by \citet{liu:2000}, which requires the calculation of the recombination abundance of N$^{+2}$ (see Tab. \ref{tab:ionic-He}). This contribution was calculated to be $< 7\%$ for the REOSC-Echelle observations.

Thus, the determination of an electronic temperature from the CELs of M\,3-27 by using the ``traditional'' method, based on the intersection of a density sensitive line diagnostic with a temperature sensitive  line diagnostic, is not possible. To estimate the $T_e$, we decided to use a methodology similar to the one presented by \citet{wesson:2005} for the analysis of this same object: by assuming a density $n_e = 10^7$ cm$^{-3}$, (as estimated from our diagnostic diagrams, as explained above), $T_e$ is determined from the [\ion{O}{iii}] $\lambda \lambda$(5007+4959)/4363 line ratio. Although it is possible to determine a $T_e$ from [\ion{O}{i}] $\lambda \lambda$(6364+6300)/5577 line ratio, this $T_e$ value was not used because it corresponds to an outer and neutral zone of the nebula where the gas is not ionized. In Fig. \ref{fig:diags} $n_e = 10^7$ cm$^{-3}$ is plotted as a vertical solid black line and by using \textsc{PyNeb} \textit{getTemden} routine the value for $T_e$ was computed. This was applied to each OAN-SPM observations and the respective $T_e$ was assumed for the whole nebula, for each observation. 

Physical conditions from optical recombination lines (ORLs) were  also determined for some of the observations. The temperature from \ion{H}{i} Balmer Jump (BJ) was only determined for the REOSC-Echelle 2019 data, following the procedure given by \citet{liu:2001}. However, the derived value is very uncertain and was not used for further calculations. 

The temperature from \ion{He}{i} was determined from the line ratio $\lambda \lambda$7281/6678 using the relation given by \citet{zhang:2005}, which is based on the \citet{benjamin:1999} atomic parameters \footnote{$T_e$(\ion{He}{i}) can be determined also using \citet{porter:2012, porter:2013} atomic data and \textsc{PyNeb}. We performed such a calculation, obtaining a temperature slightly higher that does not affect the abundance  results. Therefore, we kept the value derived with \citet{zhang:2005} method.}. $T_e$(\ion{He}{i}) could only be determined for Boller \& Chivens observations from 2004 (300 l mm$^{-1}$) and 2021 (600 l mm$^{-1}$) and REOSC-Echelle 2019, while for the other observations, the  $\lambda$7281 line was outside the wavelength range of the spectra.

The $T_e$(\ion{He}{i}) was adopted for the recombination lines of heavy elements. 
\ion{O}{ii} density was determined from the $\lambda \lambda$4649/4661 line ratio only for REOSC-Echelle 2019 spectrum.

In Table \ref{tab:phys-cond}, the values of temperature and density from CELs and ORLs are listed for the different epochs of observation. The more complete data correspond to the REOSC-Echelle 2019 observations, where electron temperatures from CELs have values from 11,100 K for [\ion{O}{i}] to 16,800 K for [\ion{O}{iii}] and electron densities go from a few  thousand cm$^{-3}$ for the outer zone to several times 10$^7$ cm$^{-3}$ in the inner zone. As expected, \ion{He}{i} temperatures, with values between $\sim 8300 - 10,500$ K, are lower than the temperatures derived from CELs.  The values listed in this table will be used to determine ionic abundances. 

\defcitealias{kingsburgh:1994}{KB94}
\defcitealias{delgado-inglada:2014b}{DI14}
\defcitealias{liu:2000}{LS00}
\defcitealias{peimbert:2014}{P14}

\begin{table*}

\caption{Electron temperatures and densities determined from CELs and ORLs for different epochs. $T_e$ in K and $n_e$ in cm$^{-3}$.  Values marked with * were determined by assuming a $n_e = 1 \times 10^7$ cm$^{-3}$. Electron densities were derived by adopting the $T_e$ determined from [\ion{O}{iii}]. :: indicates a very uncertain value.}

\begin{tabular}{lccccccc} \hline
Parameter/Obs. & Line ratio & B\&Ch 2004 (300) & Echelle 2004 & Echelle 2019 & B\&Ch 2021 (300) & B\&Ch 2021 (600) & Echelle 2021 \\ \hline \hline
 
$T_e$ [\ion{O}{i}] & (6364+6300)/5577 & $\text{15,000} \pm \text{1000}$ & $\text{11,200} \pm \text{300}$ & $\text{11,100} \pm \text{300}$ & $\text{12,200}_{-\text{1300}}^{+\text{1500}}$ & $\text{10,000}_{-\text{1340}}^{+\text{2560}}$  & $\text{11,100}\pm \text{200}$ \\

$T_e$ [\ion{Ar}{iii}]* & 7136/5192   & --- & --- & $16,400 \pm 1000$ & --- & $\text{28,420::}$ & $\text{19,800}_{-\text{1700}}^{+\text{2100}}$ \\

$T_e$ [\ion{O}{iii}]* & (5007+4959)/4363 & --- & $18,100_{-900}^{+1000}$ & $\text{16,800}_{-\text{600}}^{+\text{900}}$& $\text{16,300} \pm \text{100}$ & $\text{18,110}_{-\text{8550}}^{+\text{2550}}$  & $\text{18,000} \pm \text{800}$ \\

\textit{Adopted} & \textit{Unique zone}  & --- & $18,100_{-900}^{+1000}$ & $\text{16,800}_{-\text{600}}^{+\text{900}}$& $\text{16,300} \pm \text{100}$ & $\text{18,110}_{-\text{8550}}^{+\text{2550}}$ & $\text{18,000} \pm \text{800}$ \\ \hline

% &  &  &  &  &  &  \\

$n_e$ [\ion{N}{i}] & 5200/5198  & --- & --- & $\text{2300}_{-\text{1100}}^{+\text{4100}}$& --- & --- & $\text{700}_{-\text{400}}^{+\text{700}}$ \\

$n_e$ [\ion{O}{ii}] & 3729/3726 & --- & --- & $\text{2700}_{-\text{1900}}^{+\text{7200}}$ & --- & --- & $\text{2200}_{-\text{1500}}^{+\text{3800}}$ \\

$n_e$ [\ion{S}{ii}] & 6731/6716 & $\text{2600}_{-\text{500}}^{+\text{600}}$ & $\text{1700}_{-\text{700}} ^{+\text{1400}}$ & $\text{1400} \pm \text{400}$ & $\text{2200}_{-\text{1600}}^{+\text{5100}}$& $\text{1580}_{-\text{510}}^{+\text{440}}$  & $\text{3400}_{-\text{1000}}^{+\text{1700}}$ \\

$n_e$ [\ion{Cl}{iii}] & 5538/5518   & --- & --- & --- & $\text{4600}_{-\text{3800}}^{+\text{12,700}}$& --- & --- \\

$n_e$ [\ion{S}{ii}]b & (6716+6731)/4068 & --- & $\text{24,100}_{-\text{23,200}}^{+\text{4,500}}$ & $\text{63,300}_{-\text{62,100}}^{+\text{5400}}$ & --- & --- & $\text{0.90}_{-\text{0.88}}^{+\text{0.09}}$ \\

$n_e$ [\ion{N}{ii}] ($\times \text{10}^{\text{5}}$) & (6584+6548)/5755 & --- & $\text{5.43}_{-\text{0.97}}^{+\text{1.23}}$ & $\text{9.73}_{-\text{0.96}}^{+\text{1.12}}$ & --- & --- & $\text{12.73}_{-\text{1.17}}^{+\text{1.36}}$ \\

%$n_e$ [\ion{Ar}{iii}] ($\times \text{10}^{\text{7}}$)   & 7136/5192 & --- & --- & $\text{3.16}_{-\text{0.39}}^{+\text{0.66}}$ & --- & --- & $\text{5.39}_{-\text{1.13}}^{+\text{1.30}}$ \\

%$n_e$ [\ion{O}{iii}] ($\times \text{10}^{\text{7}}$) & (5007+4959)/4363 & $\text{3.28}_{-\text{0.81}}^{+\text{1.16}}$ & $\text{6.74}_{-\text{2.05}}^{+\text{1.69}}$ & $\text{5.55}_{-\text{0.98}}^{+\text{1.50}}$ & $\text{2.47}_{-\text{0.88}}^{+\text{2.11}}$& --- & $\text{7.84}_{-\text{1.32}}^{+\text{1.08}}$ \\

$n_e$ [\ion{Ar}{iv}]b $(\times \text{10}^{\text{7}}$) & 7170/4740   & --- & --- & $\text{5.76}_{-\text{2.29}}^{+\text{2.94}}$ & --- & --- & --- \\

\textit{Adopted}    & \textit{Outer zone} & $\text{2600}_{-\text{500}}^{+\text{600}}$ & $\text{1700}_{-\text{700}} ^{+\text{1400}}$ & $1400 \pm 400$ & $\text{2200}_{-\text{1500}}^{+\text{4900}}$ & $\text{1580}_{-\text{510}}^{+\text{440}}$ & $\text{3400}_{-\text{1000}}^{+\text{1700}}$ \\

 & \textit{Inner zone} & $1 \times 10^7$ & $1 \times 10^7$ & $1 \times 10^7$ & $1 \times 10^7$ & $1 \times 10^7$ & $1 \times 10^7$ \\ \hline

 %&  &  &  &  &  &  \\
$T_e$ \ion{H}{i} (BJ) & --- & --- & --- & 11,500:: & --- & 6280$\pm$3210 & --- \\

$T_e$ \ion{He}{i} & 7281/6678   & $\text{8500} \pm \text{400}$ & --- & $\text{8300} \pm \text{1000}$ & --- & $\text{10530}\pm  \text{4550}$ & --- \\

%$T_e$ He \textsc{i} (7281/5876) & $\text{6,100} \pm \text{200}$ & --- & $\text{7,600} \pm \text{600}$ & --- & $\text{7,200}$ & --- \\

%$T_e$ He \textsc{i} (6678/5876) & $\text{3,200} \pm \text{100}$ & $\text{6,200} \pm \text{100}$ & $\text{4,100} \pm \text{200}$ & $\text{4,000}_{-\text{1,300}}^{+\text{2,200}}$ & $\text{3,700}$ & $\text{3,200} \pm \text{100}$ \\

%$T_e$ He \textsc{i} (6678/4471)& --- & $\text{5,900}_{-\text{1,900}} ^{+\text{100}}$ & $\text{5,900} \pm \text{100}$ & $\text{5,900}_{-\text{2,600}}^{+\text{100}}$ & 3,700 & $\text{3,100} \pm \text{100}$ \\
%$T_e$ He \textsc{i} (5876/4471) & $\text{7,400} \pm \text{100}$ & $\text{2,400}_{-\text{500}} ^{+\text{600}}$ & $\text{2,300} \pm \text{100}$ & $\text{2,300}_{-\text{400}}^{+\text{800}}$ & 3,100 & $\text{3200} \pm \text{200}$ \\

%$T_e$ \ion{O}{ii} \citepalias{peimbert:2014} & --- & --- & $\text{5100}_{-\text{200}} ^{+\text{300}}$ & $\text{5200} \pm \text{100}$ & --- & --- & $\text{5200} \pm \text{100}$ \\

$n_e$ \ion{O}{ii}   & 4649/4661 & --- & --- & $\text{13,100}_{-\text{10,200}}^{+\text{12,300}}$ & --- & --- & --- \\ \hline %$\text{14,000}_{-\text{10,900}} ^{+\text{11,100}}$ 2004ec, $\text{14,100}_{-\text{11,500}} ^{+\text{11,000}}$ 2021ec

\end{tabular}

\label{tab:phys-cond}
\end{table*}

\subsection{Ionic and total abundances}

Given that  the \ion{H}{i} Balmer lines seem to be emitted partially by the star, it is inadequate to  calculate the chemical abundances following the usual procedure, with the emission line intensities relative to H$\beta$. However, considering that previous authors have calculated the chemistry in M\,3-27 by following the usual procedure we decided to calculate ionic and total abundances relative to H$^+$ and the results are shown in Appendix A. These calculations were performed using the routine \textit{get.IonAbundance} from \textsc{PyNeb}. It is clear that these results are not confident due to the contamination by the central star but we include them here for comparison and for completeness. 

In the following section  ionic abundances are calculated relative to the He$^+$.  Line intensities normalized to \ion{He}{i} $\lambda$5876, which is undoubtedly emitted by the nebula, are used. The lines used to derive ionic abundances are listed in Table \ref{tab:m3-27_lineas_ionab}. Ionic abundances relative to He$^+$ will be escalated to H$^+$ by assuming a value for the He$^+$/H$^+$ abundance ratio and in further sections, total abundances will derived.  

\begin{table}
 \centering
 \caption{\small Lines used for ionic abundances calculation.}
    \begin{tabular}{ll} \hline
    X$^{\rm{+i}}$ & Lines (\AA) \\ \hline \hline 
    N$^+$ & [\ion{N}{ii}] $\lambda\lambda$6548, 6584, 5755\\
    O$^+$ & [\ion{O}{ii}] $\lambda\lambda$3727+, 7325+ \\
    O$^{+2}$ & [\ion{O}{iii}] $\lambda\lambda$4959, 5007, 4363 \\
    Ne$^{+2}$ & [\ion{Ne}{iii}] $\lambda\lambda$3868, 3967 \\
    S$^+$ & [\ion{S}{ii}] $\lambda\lambda$6716, 6731, 4068 \\
    S$^{+2}$ & [\ion{S}{iii}] $\lambda$6312 \\
    Cl$^{+2}$ & [\ion{Cl}{iii}] $\lambda\lambda$5517, 5537 \\
    Ar$^{+2}$ & [\ion{Ar}{iii}] $\lambda\lambda$5192, 7136 \\
    Ar$^{+3}$ & [\ion{Ar}{iv}] $\lambda\lambda$4711,4740, 7170 \\
    Fe$^{+}$ & [\ion{Fe}{ii}] $\lambda$7155 \\
    Fe$^{+2}$ & [\ion{Fe}{iii}] $\lambda\lambda$4659, 4701, 4755\\
    \hline
    \end{tabular}
    \label{tab:m3-27_lineas_ionab}
\end{table}

\subsubsection{Ionic abundances relative to He$^+$}

To determine ionic abundances relative to He$^+$  we used the data of REOSC-Echelle 2019 and B\&Ch 2021 (300 l mm$^{-1}$), which are the most complete spectra of those obtained at OAN-SPM.  As said above, the line intensities were normalized to the intensity of  \ion{He}{i} $\lambda$5876. Ionic abundance of  elements relative to He$^+$ can be calculated using the next expression:

\begin{equation}
 \frac{X^{+i}}{He^{+}} = \frac{I(\lambda)}{I(\textit{\ion{He}{i}} \lambda 5876)} \frac{\epsilon_{\textit{\ion{He}{i}} \lambda 5876} (He^{+},\nu_{(\textit{\ion{He}{i}} \lambda 5876}), T{_e})}{\epsilon_{\lambda} (X^{+i},\nu_{ij}, T{_e})}
\label{eq:ionicabhe}
\end{equation}

Here $\epsilon_{\lambda}$ is the emissivity of the line used to calculate the abundance, $\epsilon_{\textit{\ion{He}{i}}\lambda5876}$ is the emissivity of the \ion{He}{i} $\lambda5876$ line derived using the physical conditions previously determined for \ion{He}{i}, and $I(\lambda)/I(\textit{\ion{He}{i}} \lambda 5876)$ is the line intensity relative to \ion{He}{i} $\lambda$5876. Line emissivities were calculated by using \textsc{PyNeb} \textit{getEmissivity} task. 

Following this method, ionic abundances from CELs relative to He$^+$ were derived by assuming a unique value for $T_e$ given by the [\ion{O}{iii}] line diagnostic and two density zones as presented in Table \ref{tab:phys-cond}.  For the once ionized species and Fe$^{+2}$ the $n_e$[\ion{S}{ii}] was used except in the case of N$^+$, for which 10$^6$ cm$^{-3}$ was adopted as derived by the line ratio [\ion{N}{ii}] $\lambda\lambda (6584+6548)/5755$. A density of 10$^7$ cm$^{-3}$ was used for the twice or more ionized species. 

Ionic abundances from the recombination lines of the ions  O$^{+2}$, N$^{+2}$ and C$^{+2}$, relative to He$^+$, were computed in the same way and by assuming the $T_e$ and $n_e$ derived from recombination lines. The $T_e$ used  was the 
$T_e$(\ion{He}{i}), while $n_e$(\ion{O}{ii}) determined from the 2019 REOSC-Echelle data was assumed in all cases. C$^{+2}$ abundance was derived from \ion{C}{ii} $\lambda$4267, O$^{+2}$ abundance was obtained from the lines of multiplet V1 and N$^{+2}$ abundance was derived from multiplet V3 lines. 

The lines used to determine ionic abundances are those listed in Table \ref{tab:m3-27_lineas_ionab}. The derived ionic abundances are presented in Table \ref{tab:ionic-He}. 

From the ionic abundances relative to He$^+$ ($X^{+i}/He^+$ in Table \ref{tab:ionic-He}), ionic abundances relative to H$^+$ can be calculated by assuming that M\,3-27 is a galactic disc PN with He/H = 0.11 \citep{kingsburgh:1994} and that He/H = He$^+$/H$^+$, that is, we assume that there is neither He$^0$ nor He$^{+2}$ present in the nebula. In this way, the ionic abundances relative to H$^+$ were calculated as: 

\begin{equation}
    X^{+i}/H^+_{(He^+)} = X^{+i}/He^+ \times He^{+}/H^+ = 0.11 \times X^{+i}/He^+ .
\end{equation}

Their values are listed in Table \ref{tab:ionic-He}.  In this table  we can appreciate that O$^{+2}$ is the most abundant ion of oxygen, being 2 or 3 orders of magnitude larger than O$^+$ calculated from $\lambda$3727+ lines (this is not the case of the O$^+$ abundance derived from the $\lambda$7325+ lines of the B\&Ch 2021 data which is only one order of magnitude lower than O$^{+2}$). For the ions [\ion{N}{ii}], [\ion{O}{iii}], [\ion{Ar}{iii}] and [\ion{Ar}{iv}] the abundances derived from the auroral and trans-auroral lines are similar indicating that used $T_e$ and densities are correct. 

The ionic abundances obtained this way can be compared to the values derived directly from H$^+$ (Appendix A). This comparison reveals that the ionic abundances from CELs, $X^{+i}/H^+_{(He^+)}$, are a factor of 1.2 lower than the values derived by using the H lines $X^{+i}/H^+$, possibly due to the assumption that $\rm{He^+/H^+} = \rm{He/H} = 0.11$. This assumption is one of those that produces large uncertainties in the abundances derived here. If we consider than He/H $=$ 0.127, the ionic abundances derived from He$^+$ and from H$^+$ would be the same. This He/H value is at $+1$ standard deviation of the mean of the distribution calculated by \citet{kingsburgh:1994} for non-Type I galactic disc PNe, which is He/H $=$ 0.112$\pm$0.015.

The ionic abundances $X^{+i}/H^+_{(He^+)}$ derived from ORLs have diminished by a larger factor $\sim 2$ when they are compared to the ionic abundances calculated directly by using H$\beta$ line, $X^{+i}/H^+$. This has consequences in the calculus of the ADF(O$^{+2}$), presented in the same Table, which diminishes from 6.04 (Appendix A) to 3.26.

\begin{table*}
\caption{Ionic abundances relative to He$^{+}$  derived from REOSC Echelle 2019 and B\&Ch 2021 data. A unique value for $T_e$ given by the [\ion{O}{iii}] line diagnosis and two density zones, as they are presented in Table \ref{tab:phys-cond}, were adopted to determine ionic abundances. Abundances relative to H$^{+}$ (columns $X^{+i}/H^+_{(He^+)}$) were determined by assuming that He/H = 0.11, as derived by \citet{kingsburgh:1994} for non-type I disc PNe, and by assuming that the whole He in M\,3-27 is single ionized (He/H = He$^+$/H$^+$).}

\begin{tabular}{lcclcc} \hline

%\multicolumn{3}{c}{$X^{+i}/He^+$} & \multicolumn{3}{c}{$X^{+i}/H^+_{(He^+)}$} \\ 
 & \multicolumn{2}{c}{$X^{+i}/He^+$} & & \multicolumn{2}{c}{$X^{+i}/H^+_{(He^+)}$}\\
 Ion  & Echelle 2019 & B\&Ch 2021 (300) & Ion  & Echelle 2019 & B\&Ch 2021 (300) \\ \hline \hline
 
ORLs &  &  &  &  &  \\ 

He$^+$ $\lambda$5876 \AA & 1 & 1 & He$^+$ $\lambda$5876 \AA & 0.11 & 0.11 \\

O$^{+2}$ $\lambda$4638 \AA \,  ($\times 10^{-3}$) & $7.87_{-2.94}^{+2.52}$ & --- & O$^{+2}$ $\lambda$4638 \AA \,    ($\times 10^{-4}$) & $8.63_{-3.23}^{+2.76}$ & --- \\

O$^{+2}$    $\lambda$4641 \AA \,  ($\times   10^{-3}$) & $4.63_{-0.57}^{+0.54}$ & --- & O$^{+2}$  $\lambda$4641 \AA \,  ($\times 10^{-4}$) & $5.08_{-0.63}^{+0.59}$ & --- \\

O$^{+2}$ $\lambda$4649 \AA \, ($\times   10^{-3}$) & $4.79_{-0.49}^{+1.00}$ & --- & O$^{+2}$ $\lambda$4649 \AA \, ($\times 10^{-4}$) & $5.25_{-0.54}^{+1.09}$ & --- \\

O$^{+2}$ $\lambda$4661 \AA  \, ($\times 10^{-3}$) & $4.40_{-1.93}^{+1.24}$ & --- & O$^{+2}$ $\lambda$4661 \AA  \,   ($\times 10^{-4}$) & $4.83_{-2.12}^{+1.36}$ & --- \\

O$^{+2}$ $\lambda$4676 \AA \, ($\times   10^{-4}$) & $5.27_{-1.45}^{+1.44}$ & --- & O$^{+2}$ $\lambda$4676 \AA \, ($\times 10^{-3}$) & $5.78_{-1.59}^{+1.58}$ & --- \\

O$^{+2}$ V1 ($\times 10^{-3}$) & $5.02_{-0.50}^{+0.46}$ & --- & O$^{+2}$ V1 ($\times 10^{-3}$) & $5.50_{-0.55}^{+0.50}$ & --- \\

C$^{+2}$ $\lambda$4267 \AA \, ($\times   10^{-4}$) & $8.90_{-3.03}^{+3.37}$ & --- & C$^{+2}$ $\lambda$4267 \AA \, ($\times 10^{-5}$) & $9.76_{-3.33}^{+3.70}$ & --- \\

N$^{+2}$ $\lambda$5679 \AA \, ($\times   10^{-3}$) & $1.69_{-0.51}^{+0.65}$ & --- & N$^{+2}$ $\lambda$5679 \AA \, ($\times 10^{-4}$) & $1.86_{-0.56}^{+0.71}$ & --- \\

N$^{+2}$ V3 ($\times 10^{-3}$) & $1.71_{-0.55}^{+0.63}$ & --- & N$^{+2}$ V3 ($\times 10^{-4}$) & $1.87_{-0.60}^{+0.69}$ & --- \\

CELs &  &  &    &  &  \\

O$^{+}$ Neb. ($\times 10^{-7}$) & $5.74_{-2.15}^{+2.79}$ & --- & O$^{+}$ Neb. ($\times 10^{-7}$) & $0.63_{-0.24}^{+0.31}$ & --- \\

O$^{+}$ Aur. ($\times 10^{-4}$) & --- & $1.46_{-0.45}^{+0.71}$ & O$^{+}$ Aur. ($\times 10^{-5}$) & --- & $1.60_{-0.49}^{+0.78}$ \\

O$^{+2}$ Neb. ($\times 10^{-3}$) & $1.53_{-0.29}^{+0.28}$ & $1.70_{-0.23}^{+0.25}$ & O$^{+2}$ Neb. ($\times 10^{-4}$) & $1.68_{-0.32}^{+0.31}$ & $1.86_{-0.25}^{+0.28}$ \\

O$^{+2}$ Aur. ($\times 10^{-3}$) & $1.53_{-0.29}^{+0.27}$ & $1.68_{-0.14}^{+0.19}$ & O$^{+2}$ Aur. ($\times 10^{-4}$) & $1.68_{-0.32}^{+0.30}$ & $1.84_{-0.15}^{+0.21}$ \\

N$^{+}$ Neb. ($\times 10^{-5}$) & $2.09 \pm 0.28$ & $2.01_{-0.23}^{+0.26}$ & N$^{+}$ Neb. ($\times 10^{-6}$) & $2.29 \pm 0.31$ & $2.20_{-0.26}^{+0.28}$ \\

N$^{+}$ Aur. ($\times 10^{-5}$) & $1.15_{-0.18}^{+0.19}$ & $2.01_{-0.24}^{+0.26}$ & N$^{+}$ Aur. ($\times 10^{-6}$) & $1.26_{-0.19}^{+0.21}$ & $2.21_{-0.26}^{+0.28}$ \\

Ne$^{+2}$ ($\times 10^{-4}$) & $3.77 \pm 0.61$ & $5.44_{-0.51}^{+0.74}$ & Ne$^{+2}$ ($\times 10^{-5}$) & $4.13 \pm 0.67$ & $5.96_{-0.56}^{+0.81}$ \\

Ar$^{+2}$ Neb. ($\times 10^{-6}$) & $6.51_{-0.94}^{+1.08}$ & $7.01_{-1.12}^{+1.17}$ & Ar$^{+2}$ Neb. ($\times 10^{-7}$) & $7.13_{-1.03}^{+1.18}$ & $7.69_{-1.23}^{+1.28}$ \\

Ar$^{+2}$ Aur. ($\times 10^{-6}$) & $6.33_{-0.99}^{+0.91}$ & --- & Ar$^{+2}$ Aur. ($\times 10^{-7}$) & $6.93_{-1.09}^{+1.00}$ & --- \\

Ar$^{+3}$ Neb. ($\times 10^{-7}$) & $7.79_{-3.04}^{+3.29}$ & --- & Ar$^{+3}$ Neb. ($\times 10^{-8}$) & $8.54_{-3.33}^{+3.60}$ & --- \\

Ar$^{+3}$ Aur. ($\times 10^{-7}$) & $15.17_{-2.77}^{+2.99}$ & --- & Ar$^{+3}$ Aur. ($\times 10^{-8}$) & $16.62_{-3.02}^{+3.28}$ & --- \\

S$^{+}$ Neb. ($\times 10^{-7}$) & $1.34_{-0.16}^{+0.21}$ & $1.88_{-0.49}^{+1.56}$ & S$^{+}$ Neb. ($\times 10^{-8}$) & $1.48_{-0.20}^{+0.18}$ & $2.06_{-0.53}^{+1.71}$ \\

S$^+$ T-aur. ($\times 10^{-6}$) & $1.40_{-0.15}^{+0.22}$ & --- & S$^+$ T-aur. ($\times 10^{-7}$) & $1.51_{-0.16}^{+0.23}$ & --- \\

S$^{+2}$ ($\times 10^{-5}$) & $1.70_{-0.24}^{+0.31}$ & $1.63_{-0.19}^{+0.24}$ & S$^{+2}$ ($\times 10^{-6}$) & $1.87_{-0.27}^{+0.34}$ & $1.79_{-0.21}^{+0.27}$ \\

Cl$^{+2}$ ($\times 10^{-5}$) & --- & $1.39 \pm 0.69$ & Cl$^{+2}$ ($\times 10^{-6}$) &  & $1.52 \pm 0.76$ \\

Fe$^{+}$ ($\times 10^{-6}$) & $1.45_{-0.30}^{+0.36}$ & --- & Fe$^{+}$ ($\times 10^{-7}$) & $1.59_{-0.33}^{+0.39}$ & --- \\

Fe$^{+2}$ ($\times 10^{-7}$) & $3.09_{-0.47}^{+0.58}$ & --- & Fe$^{+2}$ ($\times 10^{-8}$) & $3.38_{-0.52}^{+0.64}$ & --- \\
\hline
ADF(O$^{+2}$) & $3.26_{-0.61}^{+0.81}$ & --- & ADF(O$^{+2}$) & $3.26_{-0.61}^{+0.81}$ & --- \\ \hline

\end{tabular}

\label{tab:ionic-He}

\end{table*}

\subsection{Total abundances}

Total abundances of O/H, N/H, Ne/H, Ar/H, S/H, and Cl/H were computed from the {$X^{+i}/H^+_{(He^+)}$ ionic abundances (relative to He$^+$ and scaled to H$^+$)  and employing the  Ionization Correction Factors, ICFs, by \citet{kingsburgh:1994}. The ICF by  \citet{liu:2000} was used for Cl/H abundance \footnote{ICFs allow to correct total abundances for the presence of unseeing ions and the expressions used to obtain total abundance calculations are presented in Appendix \ref{section:icfs_exp}.}. The derived abundance values  are listed in Table \ref{tab:totab_he}, in units of 12+log(X/H). Also the elemental abundances relative to O are listed. The N/H values marked with :: are extremely uncertain due to the large ICFs used. 

\begin{table}

\centering

\caption{\small Total abundances of M\,3-27 from REOSC-Echelle spectrum determined using $X^{+i}/H^+_{(He^+)}$ ionic abundances. \citetalias{kingsburgh:1994} \citep{kingsburgh:1994} and \citetalias{liu:2000} \citep{liu:2000}. :: indicates a very uncertain value.}

\begin{tabular}{lcc} \hline

$12+log(X/H)$ & Echelle 2019 & B\&Ch 2021 (300) \\ \hline \hline

He/H & 11.04 & 11.04 \\

O/H & $8.22_{-0.09}^{+0.07}$ & $8.30 \pm 0.06$ \\

%N/H & 9.79::$^a$ & $7.44 \pm 0.17$ \\

N/H & 9.79:: & $7.44 \pm 0.17$ \\

ICF(N) \citepalias{kingsburgh:1994} & 2568:: & $12.32_{-4.02}^{+5.91}$ \\

Ne/H & $7.62_{-0.08}^{+0.07}$ & $7.81_{-0.04}^{+0.05}$ \\

ICF(Ne) \citepalias{kingsburgh:1994} & 1 & $1.09_{-0.03}^{+0.05}$ \\

Ar/H & $5.90_{-0.07}^{+0.06}$ & $6.16_{-0.08}^{+0.07}$ \\

ICF(Ar) \citepalias{kingsburgh:1994} & 1 & 1.87 \\

S/H & $7.26_{-0.10}^{+0.11}$ & $6.47_{-0.07}^{+0.09}$ \\

ICF(S) \citepalias{kingsburgh:1994} & $9.50^{+1.10}_{-1.58}$ & $1.65_{-0.18}^{+0.21}$ \\

Cl/H & --- & $6.40_{-0.29}^{+0.20}$ \\

ICF(Cl) \citepalias{liu:2000} & --- & $1.67_{-0.19}^{+0.22}$ \\ \hline

$log(X/O)$ &  &  \\

N/O \citepalias{kingsburgh:1994} & 1.55::& $-0.87 \pm 0.17$ \\

Ne/O \citepalias{kingsburgh:1994} & $-0.61 \pm 0.02$ & $-0.49_{-0.02}^{+0.03}$ \\

Ar/O \citepalias{kingsburgh:1994} & $-2.32 \pm 0.03$ & $-2.15 \pm 0.05$ \\

S/O \citepalias{kingsburgh:1994} & $-0.96_{-0.06}^{+0.08}$ & $-1.83 \pm 0.07$ \\

Cl/O \citepalias{liu:2000} & --- & $-1.89_{-0.29}^{+0.17}$ \\ \hline
%\multicolumn{3}{l}{$^a$ :: indicate a very large uncertainty for N abundance.}

\end{tabular}

\label{tab:totab_he}

\end{table}

\section{Expansion velocities from lines}

Kinematic analysis from CELs and ORLs was performed for the data from REOSC-Echelle spectra, as these data offer the best spectral resolution (R $\sim$ 18,000) and cover a large range of wavelengths. For this analysis, the FWHMs of lines were measured and corrected by the effect of thermal and instrumental broadening. For the correction by thermal broadening the electron temperature adopted for each observation (as presented in Table \ref{tab:phys-cond}) was used.

The FWHMs, expansion velocities $v_{exp}$ (determined from FWHM of lines) and radial velocities $v_{rad}$  of the observed lines in the three REOSC-Echelle spectra (2004, 2019 and 2021) are presented as online material together with the line fluxes and intensities for these observations. See an example of these data in  Table \ref{tab:line-intensities}.  In Fig. \ref{fig:kinematics} (left) we present the diagram of expansion velocities in function of ionization potentials for different CELs and ORLs (CELs represented by blue points, ORLs by red diamonds, green points correspond to CEL aurorals); it was constructed using REOSC-Echelle data from 2019 because this spectrum presents the largest number of measured lines. This diagram is equivalent to plot $v_{exp}$  vs. distance to the central star, as it is expected that, due to the ionization structure of the nebula, ions more highly ionized are located nearer the central star \citep[see e.g., the work by][and references therein]{pena:2017}.

\begin{figure*}
	\includegraphics[width=\columnwidth]{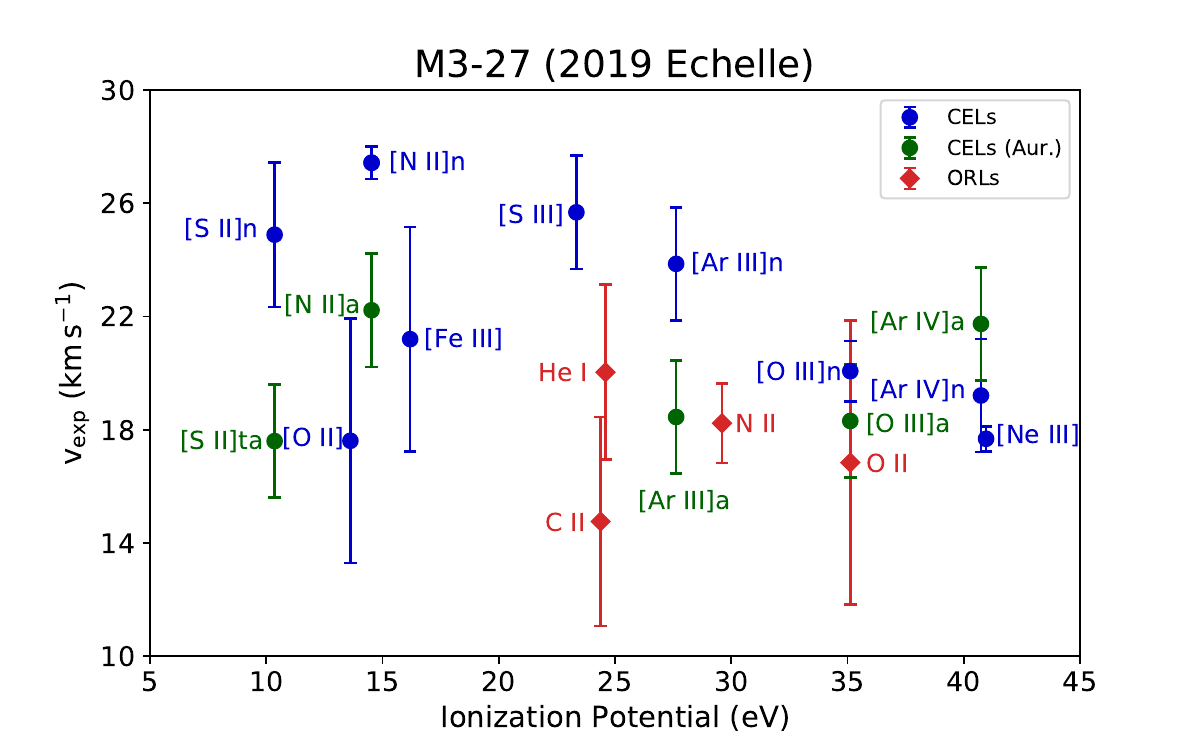}
	\includegraphics[width=\columnwidth]{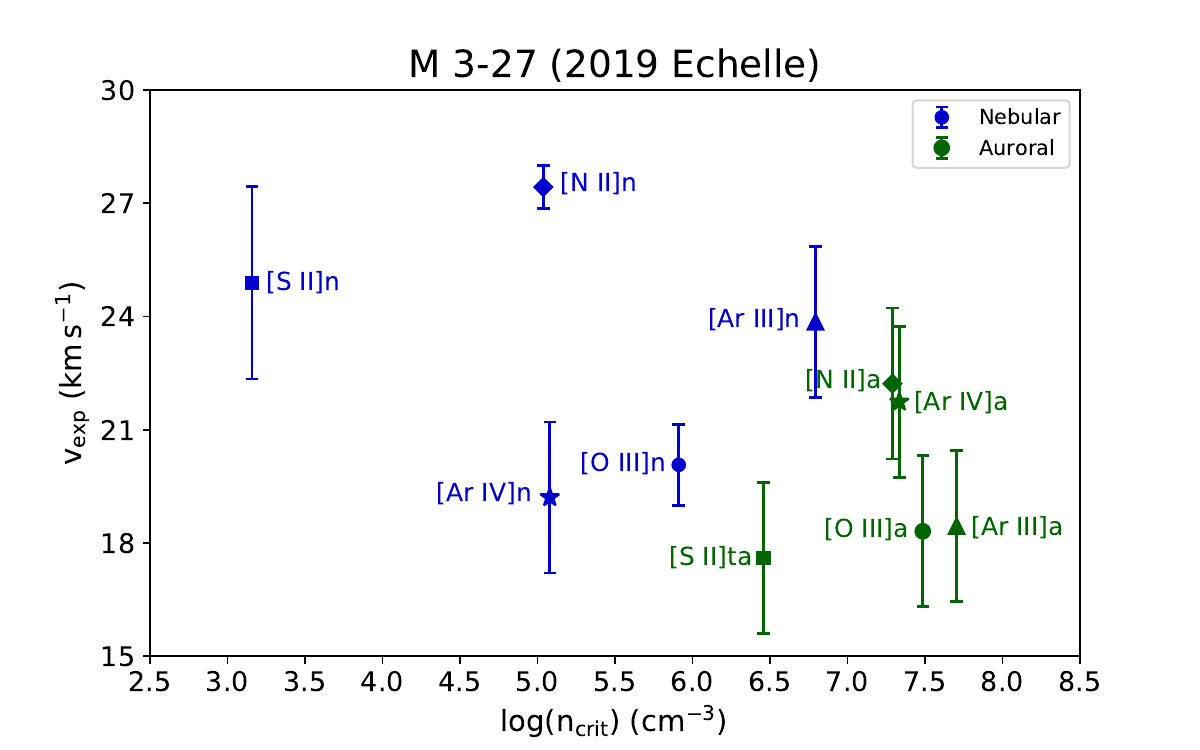}
    \caption{(Left) Expansion velocities of ions as a function of their ionization potentials for our REOSC-Echelle observations. $v_{exp}$ of CELs are in blue (nebular lines) or green (auroral and trans-auroral lines), $v_{exp}$ from ORLs are in red. Right figure: Expansion velocities of nebular (blue), auroral and trans-auroral lines (in green) vs. critical densities.}
    \label{fig:kinematics}
\end{figure*}

In Fig. \ref{fig:kinematics} (left) it is observed that lines of [\ion{S}{ii}], [\ion{N}{ii}] and [\ion{S}{iii}] show velocities larger than 25 km s$^{-1}$, while lines of [\ion{Fe}{iii}] and [\ion{O}{ii}] show smaller $v_{exp}$ but have large error bars. Lines of [\ion{Ar}{iii}] and [\ion{O}{iii}] present $v_{exp}$ of 25 and 21 km s$^{-1}$ and ions of larger ionization potentials like [\ion{Ne}{iii}] and [\ion{Ar}{iv}] have the lowest $v_{exp}$ of 20 and 18 km s$^{-1}$ respectively, thus CELs present a decreasing $v_{exp}$, from 26 to 18 km s$^{-1}$, with increasing ionization potential. On the other hand, recombination lines show lower $v_{exp}$ from about 15 to 19 km s$^{-1}$ which could be interpreted as if recombination lines are originated in a different zone than CELs, however  the uncertainties are large, therefore the differences in velocities are not conclusive. 

Also in Fig. \ref{fig:kinematics}, the green dots represent the $v_{exp}$ of CELs auroral and transauroral lines of [\ion{S}{ii}], [\ion{N}{ii}], [\ion{O}{iii}], [\ion{Ar}{iii}] and [\ion{Ar}{iv}], which show a different kinematic behaviour between both lines. It is observed that auroral (transauroral) lines of  [\ion{S}{ii}], [\ion{N}{ii}] and [\ion{Ar}{iii}] show lower $v_{exp}$ than their respective nebular lines, by about 7, 6 and 6 km s$^{-1}$, respectively; however, for auroral and nebular lines of  [\ion{O}{iii}] and [\ion{Ar}{iv}] this difference is within the errors bars. This difference can be attributed to the suppression of nebular lines emission due to $n_{crit}$ of these lines is lower than the $n_e$, which,  as it was shown in previous sections, increases in inner parts of the nebula. In this way, in Fig. \ref{fig:kinematics} (right) the differences in $v_{exp}$ of nebular, auroral, and transauroral lines of the same ion are presented as a function of their critical densities. It is clear that lines with higher critical densities show lower $v_{exp}$.

\section{Discussion}

The young PN M\,3-27 presents several peculiarities that make difficult to perform a classical nebular analysis. The first problem is that the \ion{H}{i} lines show P-Cygni type profiles with wide wings, these profiles are different in shape and in kinematic than the \ion{He}{i} and heavy element profiles. This indicates that the \ion{H}{i} lines are mainly emitted by the central star. We found that H$\alpha$ intensity has increased importantly between 1974 to 2004, suggesting important changes in the stellar emission in a period of 50 years. 

In addition, several nebular line intensities have changed in the same period. To follow this evolution, the line intensities relative to H$\beta$ were analysed (see Fig. \ref{fig:evolution}), but considering that H$\beta$ is mainly emitted by the central star and might be changing,  the line intensities relative to \ion{He}{i} $\lambda$5876 (which has a nebular profile) were also analysed (see Fig. \ref{fig:evolution_he}). It is found that the line intensities of [\ion{O}{iii}]$\lambda$5007, [\ion{N}{ii}]$\lambda$6584 and [\ion{O}{ii}]$\lambda$3727+ have decreased significantly, while the auroral lines [\ion{O}{iii}]$\lambda$4363, [\ion{N}{ii}]$\lambda$5755, [\ion{O}{ii}]$\lambda$7325+ and also [\ion{Ne}{iii}]$\lambda$3869 present some small variations without important changes. This can be interpreted as a suppression of some nebular lines due to the large electron density in the inner nebular zone, with a value of about 10$^7$ cm$^{-3}$ which is in the limit or above the critical density of these lines. This value of $n_e$, estimated by the [\ion{O}{iii}] $\lambda\lambda (5007+4959)/4363$ line ratio is one order of magnitude larger than the value reported in the works by \citet{adams:1975, ahern:1978, barker:1978} and \citet{kohoutek:1968} of 10$^6$ cm$^{-3}$, estimated using the same line ratio.

We do not attribute the changes of line intensities to recombination of the nebula because we do not find neither the diminution of the line intensities of high ionization species, like [\ion{Ne}{iii}]$\lambda$3869 or the auroral line [\ion{O}{iii}]$\lambda$4363, nor the increase of intensities of low ionization species lines, as [\ion{N}{ii}]$\lambda$5755 or [\ion{O}{ii}]$\lambda$7325+. Instead, the suppression of intensities is observed in lines having the lower critical densities.

Because of the details presented by \ion{H}{i} lines, to determine the interstellar logarithmic reddening, c(H$\beta$), we used a procedure based on the \ion{He}{i} lines which provides a value between  0.39  and 0.63 in agreement with values derived by other authors from data of 1970 and calculated considering self-absorption, $\sim 0.50 - 0.75$. It is interesting to note that \citet{schlegel:1998} and \citet{schlafly:2011} have reported that the interstellar extinction in M\,3-27 direction is  c(H$\beta$)=0.31, therefore a part of the value derived for M\,3-27 may be intrinsic in the nebula. 

The estimates of electron density and temperature from CELs are complicated by the fact that density present a large contrast, with a value of about 10$^3$ cm$^{-3}$ in the outer zone, and about 10$^7$ cm$^{-3}$ in the inner zone. This high inner value causes that the line ratios usually employed to determine electron temperatures  are now sensitive to density. With this density values of temperature between 16,000 and 18,000 K are found for our data. We adopted these  values to derived ionic abundances of these observations from the CELs. It should be noted that \citet{wesson:2005} derived $T_e = 13,000$ K assuming the same density for the inner zone. This occurred  because the [\ion{O}{iii}]$\lambda$4363 intensity used by them (obtained in the year 2001) is lower than our value and the values reported for the seventies.

The electron temperature can also be determined from the recombination lines of \ion{He}{i}, obtaining values of $8300 - 10,500$ K. Only for REOSC-Echelle 2019 data a temperature from the \ion{H}{i} Balmer continuum with a value of 11,500 K was derived, nevertheless, this value is very uncertain and was not used for other calculations. It was found that the temperatures from ORLs are lower than the temperature from CELs, which is an expected result for planetary nebulae \citep[this was first noticed by][]{peimbert:1971}. 

Another difficulty in M\,3-27 analysis is the determination of total abundances from CELs. Even though the determination of total O/H abundance is simple (just by adding O$^+$ and O$^{+2}$ ionic abundances) due to O$^+$ abundance from [\ion{O}{ii}]$\lambda$3727+ lines is 3 or 4 orders of magnitude smaller than O$^{+2}$ abundance (see Table \ref{tab:ionic-abundances}), ICFs for other elements that depend on O$^+$, such as N and S, will be severely affected. For REOSC-Echelle spectra, this effect is evident for the ICF(N) by \citet{kingsburgh:1994}, which would be extremely elevated ($>500$) and therefore N/H total abundance would be very uncertain and overestimated. An alternative is to determine this ICF by using the auroral lines [\ion{O}{ii}] $\lambda$7325+, as it was done for the B\&Ch spectra, in such a case the N/H abundance behaves accordingly to values of disc PNe. However, it is necessary to be cautious in the use of the N abundance derived using [\ion{O}{ii}] $\lambda$7325+ lines because, as it is well known, ionic abundances determined using this  auroral line of O$^+$ are systematically larger than the derived from the nebular line \citep[see e.g.,][]{rodriguez:2020}. Along with this problem is the diminution of O$^{+}$ nebular line intensity in a short period of time. For ICFs for Ar and Ne by \citet{kingsburgh:1994}, their values are 1 for REOSC-Echelle observations, because they are based on O$^{+2}$/O rates; in the case of Ar, in REOSC-Echelle spectra Ar$^{+3}$ is detected and so the correction is made considering the presence of this ion; for Boller \& Chivens spectra this ion is not detected and, therefore, the correction was made using the ICF by \citet{kingsburgh:1994} that considers only the Ar$^{+2}$ abundance.
\bigskip

The C$^{+2}/\rm{H}^+$  abundance derived from the recombination line $\lambda$4267 is similar to the value reported by \citet{wesson:2005} from the same recombination line. These authors also derived C$^{+2}$/H$^+$ from the ultraviolet line $\lambda$1909 intensity and adopting $T_e =13,000$ K, their derived value is similar to their value from $\lambda$4267 and would suggest an ADF(C$^{+2}$) near 1. Previously \citet{feibelman:1985} had estimated the C$^{+2}$/H$^+$ value from the same line [\ion{C}{iii}]$\lambda$1909 but adopting $T_e = 15,000$ K. Both C$^{+2}$/H$^+$ values are very different due to the large dependence of CELs in temperature. Feibelman's value is more confident and would imply an ADF(C$^{+2}$) of 1.86, more in agreement to the value derived for O$^{+2}$ and other elements.

\bigskip
\bigskip

Considering that \ion{H}{i} lines seems to be emitted by the star, we have calculated the ionic abundances relative to the He$^+$ for the observations of REOSC-Echelle 2019 and B\&Ch 2021, and transformed them to abundances relative to H$^+$ by assuming that He/H = He$^+$/H$^+$ = 0.11. When possible, the ionic abundances derived from the auroral and nebular lines were computed to compare their values.

With this procedure the chemical abundances derived from CELs diminish by a factor of 1.2, while the chemical abundances derived from ORLs diminish by a factor of 2, when compared with the abundances derived relative to H$^+$ (see Appendix A). This effect could be due to the assumptions of $T_e$ and $n_e$ made to determine the emissivity of He$^+$ and the assumption made considering that He$^+$/H$^+$ = He/H, that is, we consider that there is no He$^0$ in the nebula. To determine the contribution He$^0$ to He/H abundance is problematic and still is an open problem \citep[see the discussion presented by][]{delgado-inglada:2014b}.
This decrease in abundances has effects on the determination of the ADF(O$^{+2}$) which is 7.30 when abundances relative to H$^+$ are used  (the average value from REOSC-Echelle spectra) and 3.26 if abundances relative He$^+$ are used (from REOSC-Echelle 2019 spectrum). 

From the physical conditions and abundances derived from CELs and ORLs there are indications of the existence of two different plasmas in M\,3-27 although the kinematics of CELs and ORLs are not so conclusive in this sense.

Our results so far are affected by the complexity of the abundance analysis in M\,3-27. We consider that the best values of ionic abundances are those estimated relative to He$^+$ despite the several assumptions that were done for the He$^+$ abundance. Because of the problems for the ICFs that are based in O$^+$ abundance, we consider that the best values for total abundances are those derived with B\&Ch 2021 data. 

M\,3-27 shares similar characteristics with the also young and compact PNe Vy\,2-2 and IC\,4997 that were analysed in our previous work \citep{ruiz-escobedo:2022}. These three PNe present a large density gradient showing an outer low-density zone with $n_e \sim 10^3 - 10^4$ cm$^{-3}$ and an inner high-density zone with $n_e \sim 10^5 - 10^7$ cm$^{-3}$, being M\,3-27 the nebula with the largest gradient. The three objects have sub-solar metallicities and show differences between the $v_{exp}$ of nebular and auroral (or trans-auroral) lines of CELs of ions like [\ion{S}{ii}], [\ion{N}{ii}], [\ion{O}{iii}], [\ion{Ar}{iii}] and [\ion{Ar}{iv}]. It is very possible that, as it was suggested by \citet{zhang:2002}, the standard temperature diagnostic ratios trend to overestimate $T_e$  of the dense planetary nebulae resulting in an underestimation of total abundances.

Another important similarity between M\,3-27 and IC\,4997 is the variability of their line intensities and their  physical conditions in short periods of time.   \citet{kostyakova:2009} monitored the spectral evolution of IC\,4997 for forty years. They claimed that $n_e$ increased by a factor of 5, and $T_e$ increased from 12,000 K to 14,000 K in a period of 20 yr while the nebular ionization degree has been growing with time. The central star seems to have increased its effective temperature from $37,000 - 40,000$ K to 47,000 K in the same period, thus the nebular variations could be produced by changes in the central star which  has been heating with time. IC\,4997 also present a wide H$\alpha$ line with a FWZI equivalent to 5,100 km s$^{-1}$ attributed to Raman scattering  \citep{arrieta:2003}. In addition \citet{hyung:1994}, from $1990 - 1991$ observations, found that H$\alpha$ exhibits a P-Cygni profile. Nonetheless, \citet{miranda:2022} showed that during 2019 and 2020, the H$\alpha$ profile exhibited only a single-peak profile and that its broad wings narrowed by a factor of 2. 

Due to the above exposed, we consider that M\,3-27 is an object very similar to IC\,4997, in a previous evolutionary stage.

\section{Conclusions}

From high- and medium-spectral resolution obtained at OAN-SPM, M\'exico, from 2004 to 2021, we have analysed the characteristics of the young and dense PN M\,3-27. During this epoch the central star shows profiles in emission type P-Cygni  in the \ion{H}{i} Balmer lines, in particular in H$\alpha$, H$\beta$, H$\gamma$ and H$\delta$, indicating the presence of a stellar wind. The stellar emission is variable as deduced mainly from the H$\alpha$ emission.  Due to the stellar emission of the Balmer lines, they cannot be used for determining c(H$\beta$) from the Balmer decrement, alternatively  the \ion{He}{i} lines were used and a value of about 0.47 was obtained (see Tab. \ref{tab:c(Hbeta)}).

M\,3-27 is a very dense PN with an inner zone with density about 10$^7$ cm$^{-3}$, which is larger than several critical densities of some nebular lines. This make difficult to determine electron temperatures with the usual  nebular-auroral line intensity ratios. By adopting a density of 10$^7$ cm$^{-3}$ as derived from our diagnostic diagrams, a $T_e$ between $16,300 - 18,200$ K was derived from CELs and employed for the calculus of ionic abundances. 
The inner density value seems to have increased in a period of 30 years from 10$^6$ cm$^{-3}$  in the seventies \citep{adams:1975, ahern:1978, barker:1978,kohoutek:1968} to 10$^7$ cm$^{-3}$, value found in this work. 

Total abundances in M\,3-27 were derived by calculating ionic abundances relative to He$^+$ scalated to H$^+$,  and by using the ICFs by \citet{kingsburgh:1994} and \citet{liu:2000}. The values obtained for 12+log(O/H) are sub-solar,  between 8.23 and 8.30 (the solar value is $\text{12+log(O/H)} = 8.69$ as derived by \citealt{asplund:2009}). Our derived values for 12+log(O/H) are smaller than the value presented by \citet{wesson:2005}, probably due to the lower  temperature used by these authors for abundance determinations. Total abundances of elements derived using ICFs based on O$^{+}$ nebular line $\lambda$3727+ seem to be overestimated, due to the suppression of this line, so we adopted the total values determined using the auroral line $\lambda$7325+. The abundance values of other elements relative to O indicate that M\,3-27 is a disc nebula \citep{kingsburgh:1994}.

We have analysed the characteristics of M\,3-27 in comparison with the also young and compact PNe Vy\,2-2 and IC\,4997 which  also present a large density gradient and variability of their line intensities. In particular M\,3-27 is very similar to IC\,4997 which has also  showed stellar P-Cygni profile, nebular variations and wide H$\alpha$ wings  attributable to Raman scattering.

In the case of M\,3-27, the important variations found on the H$\alpha$ intensity suggest important changes on the central star emission. The variation of H$\alpha$ profile, from the single-peaked profile reported by \citet{tamura:1990} to the P-Cygni profile, first reported by \citet{miranda:1997} and also found in our work, suggest that the stellar wind has strengthened. This changes in the stellar emission are reflected in the nebula, where it is found an increase in $n_e$ from $10^6$ cm$^{-3}$ (in the seventies) to $10^7$ cm$^{-3}$ (after the year 2000) and the consequent suppression on the emission of lines with $n_{crit}$ lower than the $n_e$ of the nebula. Probably the stellar temperature has been changing with time too, however, this parameter was not explored in this work. 

Forthcoming observations are needed to analyse the evolution of line intensities and nebular and stellar parameters of this very young planetary nebula, which has showed important changes in its characteristics in short periods of time.

\section*{Acknowledgements}

This work received partial support from Direcci\'on General de Asuntos del Personal Acad\'emico-Programa de Apoyo a Proyectos de Investigaci\'on e Innovaci\'on Tecnol\'ogica (DGAPA-PAPIIT) IN111423, IN105020 and IN103519 and Consejo Nacional de Ciencia y Tecnolog\'ia (CONACyT), M\'exico, project A1-S-15140. FR-E acknowledges scholarship by CONACyT, M\'exico. Helpful discussion and suggestions by Dr. Michael G. Richer are deeply acknowledged. 

This work is based upon observations carried out at the Observatorio Astron\'omico Nacional at the Sierra San Pedro M\'artir (OAN-SPM), Baja California, M\'exico. We thank the daytime and night support staff at the OAN-SPM for facilitating and helping to obtain our observations. The observations during 2021 were obtained by the resident astronomers at OAN-SPM as observers were not allowed at the observatory due to the COVID19 pandemic, so we are particularly grateful to them for their effort. 

This work has made use of data from the European Space Agency (ESA) mission {\it Gaia} (\url{https://www.cosmos.esa.int/gaia}), processed by the {\it Gaia} Data Processing and Analysis Consortium (DPAC, \url{https://www.cosmos.esa.int/web/gaia/dpac/consortium}). Funding for the DPAC has been provided by national institutions, in particular the institutions participating in the {\it Gaia} Multilateral Agreement.

%%%%%%%%%%%%%%%%%%%%%%%%%%%%%%%%%%%%%%%%%%%%%%%%%%
\section*{Data Availability}

The data underlying this article will be shared on reasonable request to the corresponding author.
 
%The inclusion of a Data Availability Statement is a requirement for articles published in MNRAS. Data Availability Statements provide a standardised format for readers to understand the availability of data underlying the research results described in the article. The statement may refer to original data generated in the course of the study or to third-party data analysed in the article. The statement should describe and provide means of access, where possible, by linking to the data or providing the required accession numbers for the relevant databases or DOIs.



\bibliographystyle{mnras}
\bibliography{references_m3-27} % if your bibtex file is called example.bib

%%%%%%%%%%%%%%%%%%%%%%%%%%%%%%%%%%%%%%%%%%%%%%%%%%

%%%%%%%%%%%%%%%%% APPENDICES %%%%%%%%%%%%%%%%%%%%%

\appendix

\section{Ionic and total abundances relative to H}
From the line intensities relative to H$\beta = 100$ which are listed  in Table \ref{tab:line-intensities} ionic abundances relative to H$^+$ were determined for all the epochs of observations.
As was explained in Section 5.1, the routine \textit{get.IonAbundance} from \textsc{PyNeb} was used. The procedure is the same as in Section 5.1, a unique value for $T_e$ given by the [\ion{O}{iii}] line diagnostic and two density zones as  presented in Table \ref{tab:phys-cond} were adopted to determine ionic abundances. For the once ionized species and Fe$^{+2}$ the $n_e$[\ion{S}{ii}] was used except in the case of N$^+$ for which 10$^6$ cm$^{-3}$ was adopted. A density of 10$^7$ cm$^{-3}$ was used for the twice or more ionized species. Again when possible, the ionic abundances derived from the auroral and nebular lines were computed to compare their values.  The lines employed are those listed in Table \ref{tab:m3-27_lineas_ionab}. The derived values of ionic abundances are presented in Table \ref{tab:ionic-abundances}. 

As we concluded in Section 5.1, in Table \ref{tab:ionic-abundances} we can also appreciate that O$^{+2}$ is the most abundant ion of oxygen, being 2 or 3 orders of magnitude larger than O$^+$ calculated from $\lambda$3727+ lines and only one order of magnitude if  O$^+$ abundance is derived from the  $\lambda$7325+ lines of the B\&Ch 2021 data. And again for the ions [\ion{N}{ii}], [\ion{O}{iii}], [\ion{Ar}{iii}] and [\ion{Ar}{iv}] the abundances derived from the auroral and trans-auroral lines are similar indicating that $T_e$ and densities used are correct. 

Considering all the above, we adopted the abundances of S$^+$, N$^+$, O$^{+2}$, Ar$^{+2}$ and Ar$^{+3}$ derived from the nebular lines except in the case of N$^+$ from B\&Ch, for which the abundance derived from $\lambda$5755 was adopted\footnote{The correction of the intensity of the auroral line [\ion{N}{ii}] $\lambda$5755 was presented in Section \ref{sec:physical_conditions}. The value of the recombination contribution was $< 7$\%.
%The N$^{+2}$ abundance was corrected by subtracting the recombination contribution of the line [\ion{N}{ii}] $\lambda$5755 following the methodology by \citet{liu:2000}. The value of the recombination contribution was 7\%
}.

In general, the ionic abundances relative to H$^+$ derived from both REOSC-Echelle and B\&Ch spectra  are similar. 

\subsubsection{Abundances from ORLs}
He$^+$ abundance was derived from the line \ion{He}{i} \,$\lambda$5876. $n_e$ derived from \ion{O}{ii} $\lambda\lambda$4649/4661 determined from REOSC-Echelle 2019 spectrum was assumed for all observations. $T_e$(\ion{He}{i}) as presented in Table \ref{tab:phys-cond} was used for these calculations. For the REOSC-Echelle 2004 observation the value determined from B\&Ch (300 l mm$^{-1}$) observation from the same year was assumed, while for REOSC-Echelle and B\&Ch (300 l mm$^{-1}$) 2021 observations, the $T_e$(\ion{He}{i}) from REOSC-Echelle 2019 was assumed. All the values are presented in Table \ref{tab:ionic-abundances}.

ORLs abundances of ions of heavy elements, O$^{+2}$, N$^{+2}$ and C$^{+2}$, were only determined from REOSC-Echelle spectra. As made explicit in \S 5.1 the $T_e$ used for all cases was the $T_e$(\ion{He}{i}) determined following \citet{zhang:2005} methodology, while $n_e$(\ion{O}{ii}) determined from the 2019 REOSC-Echelle data was assumed for all cases. C$^{+2}$ abundance was derived from \ion{C}{ii} $\lambda$4267, O$^{+2}$ abundance was obtained from the lines of multiplet V1 and N$^{+2}$ abundance was derived from multiplet V3 lines. All these ionic abundances are also listed in Table \ref{tab:ionic-abundances}.

The average ADFs(O$^{+2}$) derived from these calculation is 7.37, similar to the ADF derived by \citet{wesson:2005} and  larger than the ADF obtained by using He$^+$ in the determination of ionic abundances.

\begin{table*}

\caption{Ionic abundances relative to H$^+$ for all epochs. A unique value for $T_e$, given by the [\ion{O}{iii}] line diagnostic and two density zones as they are presented in Table \ref{tab:phys-cond} were adopted to determine ionic abundances.}

\begin{tabular}{lcccccc} \hline

$X^{+i}/H^+$ &  Echelle 2004 & Echelle 2019 & B\&Ch 2021 (300) & B\&Ch 2021 (600) & Echelle 2021 \\ \hline

ORLs &    &  &  &  &  \\

He$^+$ (5876) &   $0.230_{-0.022}^{+0.020}$ & $0.241^{+0.008}_{-0.007}$ & $0.199_{-0.031}^{+0.039}$ & $0.176\pm0.050$ & $0.188_{-0.004}^{+0.005}$ \\

O$^{+2} \times 10^{-4}$ (4638) &  --- & $1.37_{-0.39}^{+0.64}$ & --- & --- & --- \\

O$^{+2} \times 10^{-4}$ (4641) &  --- & $1.12 \pm 0.13$ & --- & --- & $1.10_{-0.22}^{+0.23}$ \\

O$^{+2} \times 10^{-4}$ (4649) &  $1.35_{-0.42}^{+0.59}$ & $1.14_{-0.12}^{+0.29}$ & --- & --- & $1.17_{-0.17}^{+0.21}$ \\

O$^{+2} \times 10^{-4}$ (4650) & --- & --- & --- & --- & $1.12_{-0.56}^{+0.46}$ \\

O$^{+2} \times 10^{-4}$ (4661) &  --- & $1.06_{-0.41}^{+0.29}$ & --- & --- & $0.92_{-0.42}^{+0.36}$ \\

O$^{+2} \times 10^{-4}$ (4676) &  --- & $1.26_{-0.33}^{+0.36}$ & --- & --- & $1.89_{-0.45}^{+0.46}$ \\

O$^{+2} \times 10^{-4}$ (V1) &  $1.21_{-0.09}^{+0.12}$ & $1.22_{-0.12}^{+0.11}$ & --- & --- & $1.17_{-0.15}^{+0.14}$ \\

C$^{+2} \times 10^{-4}$ (4267) &  --- & $2.17_{-0.78}^{+0.80}$ & --- & --- & $2.25_{-0.66}^{+0.68}$ \\

N$^{+2} \times 10^{-4}$ (5679) &  $6.41_{-1.85}^{+2.24}$ & $4.06_{-1.21}^{-1.59}$ & --- & --- & $2.14_{-0.76}^{+0.95}$ \\

N$^{+2} \times 10^{-4}$ (V3) &  $6.05_{-1.88}^{+1.72}$ & $3.84_{-1.18}^{+1.37}$ & --- & --- & $2.05_{-0.72}^{+0.81}$ \\ \hline
 
CELs &  &  &  &  &  &  \\ 

O$^{+} \times 10^{-7}$ (Neb.) &  $2.57_{-1.58}^{+1.65}$ & $0.73_{-0.29}^{+0.37}$ & --- & $6.56\pm 4.40$ & $1.42_{-0.51}^{+0.63}$ \\

O$^{+} \times 10^{-6}$ (Aur.) &  --- & --- & $18.90_{-6.33}^{+8.44}$ & $15.59\pm 1.65$ & --- \\

O$^{+2} \times 10^{-4}$ (Neb.) &  $1.53_{-0.30}^{+0.36}$ & $1.99 \pm 0.39$ & $2.14 \pm 0.21$ & $1.55 \pm 0.21$ & $1.58_{-0.31}^{+0.34}$ \\

O$^{+2} \times 10^{-4}$ (Aur.) &  $1.52_{-0.30}^{+0.36}$ & $1.99_{-0.39}^{+0.37}$ & $2.16 \pm 0.06$ & $1.47 \pm 0.37$ & $1.59_{-0.29}^{+0.35}$ \\

N$^{+} \times 10^{-6}$ (Neb.) &  $2.45 _{-0.48} ^{+0.47}$ & $2.70_{-0.37}^{+0.36}$ & $2.54 \pm 0.19$ & --- & $1.80_{-0.22}^{+0.25}$ \\

N$^{+} \times 10^{-6}$ (Aur.) &  $1.23 \pm 0.22$ & $1.39_{-0.22}^{+0.21}$ & $2.54_{-0.18}^{+0.19}$ & $3.65 \pm 0.81$ & $0.97 \pm 0.16$ \\

Ne$^{+2} \times 10^{-5}$ &  $2.86 _{-0.52} ^{+0.51}$ & $4.92_{-0.77}^{+0.82}$ & $5.70 \pm 0.29$ & $4.28 \pm 0.11$ & $4.15_{-0.68}^{+0.73}$ \\

Ar$^{+2} \times 10^{-6}$ (Neb.) & --- & $0.85_{-0.13}^{+0.11}$ & $0.90 \pm 0.11$ & $0.67\pm 0.15$ & $0.57_{-0.09}^{+0.08}$ \\

Ar$^{+2} \times 10^{-6}$ (Aur.) &  $0.61_{-0.13}^{+0.14}$ & $0.82 \pm 0.12$ & --- & --- & $0.62\pm0.10$ \\

Ar$^{+3} \times 10^{-7}$ (Neb.) &  --- & $10.04_{-3.95}^{+4.07}$ & --- & $12.16\pm 0.12$ & $9.05_{-1.94}^{+1.90}$ \\

Ar$^{+3} \times 10^{-7}$ (Aur.) &  --- & $19.68_{-3.29}^{+3.84}$ & --- & --- & --- \\

S$^{+} \times 10^{-8}$ (Neb.) &  $3.74_{-0.60}^{+0.83}$ & $1.73_{-0.18}^{+0.20}$ & $2.40_{-0.63}^{+1.77}$ & $2.45\pm 1.21$ & $1.50_{-0.22}^{+0.32}$ \\

S$^{+} \times 10^{-7}$ (Taur.) &  $1.58_{-0.22}^{+0.27}$ & $1.81 \pm 0.20$ & --- & --- & $1.63_{-0.18}^{+0.16}$ \\

S$^{+2} \times 10^{-6}$ &  $1.56 \pm 0.29$ & $2.26_{-0.35}^{+0.36}$ & $2.16_{-0.16}^{+0.21}$ & $1.26\pm 0.50$ & $1.32_{-0.23}^{+0.22}$ \\

Cl$^{+2} \times 10^{-6}$ &  --- & --- & $4.36_{-1.88}^{+2.40}$ & --- & --- \\

Fe$^{+} \times 10^{-7}$ &  --- & $1.86_{-0.36}^{+0.43}$ & --- & --- & --- \\

Fe$^{+2} \times 10^{-8}$ &  --- & $4.02_{-0.67}^{+0.64}$ & --- & --- & $2.48_{-0.49}^{+0.53}$ \\ \hline

ADF(O$^{+2}$) V1 & $8.40_{-2.72}^{+3.90}$ & $6.04_{-1.08}^{+1.69}$ & --- & --- & $7.47_{-1.63}^{+1.86}$ \\ \hline

\end{tabular}

\label{tab:ionic-abundances}
\end{table*}

\begin{table*}

\caption{Total abundances in 12+log(X/H) scale, determined for all the epochs. Relative abundances to O are presented in log(X/O) scale. \citetalias{kingsburgh:1994} \citep{kingsburgh:1994} and \citetalias{liu:2000} \citep{liu:2000}. :: indicates a very uncertain value.}

\begin{tabular}{lcccccc} \hline

 &  Echelle 2004 & Echelle 2019 & B\&Ch 2021 (300) & B\&Ch 2021 (600) & Echelle 2021 \\ \hline

\textit{12+log(X/H)} &     &   &    &   & \\

He/H &  $11.36 \pm 0.04$ & $11.38 \pm 0.01$ & $11.30_{-0.07}^{+0.08}$& $11.28 \pm 0.44$ & $11.27 \pm 0.01$ \\

O/H &  $8.18_{-0.09}^{+0.10}$ & $8.30_{-0.10}^{+0.08}$ & $8.37_{-0.05}^{+0.04}$& $8.28 \pm 0.94$ & $8.20_{-0.09}^{+0.08}$ \\

N/H  \citepalias{kingsburgh:1994} &  $9.16$::  & $9.86$::  & $7.50 \pm 0.17$& $7.58 \pm 0.82$ & 9.29:: \\

ICF(N) \citepalias{kingsburgh:1994} &  575::  & 2603::   & $12.37_{-4.04}^{+5.84}$  & $1.13\pm 0.76$ & 1100::  \\

Ne/H  &  $7.46_{-0.09}^{+0.07}$ & $7.69 \pm 0.07$ & $7.79 \pm 0.02$ & $7.37 \pm 0.65$ & $7.62_{-0.08}^{+0.07}$ \\

ICF(Ne) \citepalias{kingsburgh:1994} &  1.00  &  1.00 & $1.09_{-0.03}^{+0.05}$  &  $1.05\pm 0.05$ & 1.00 \\

Ar/H &  $6.06_{-0.10}^{+0.09}$ & $5.98_{-0.07}^{+0.06}$ & $6.23_{-0.06}^{+0.05}$& $5.51 \pm 0.42$ & $5.82_{-0.07}^{+0.06}$ \\

ICF(Ar) \citepalias{kingsburgh:1994}    &  1.87  & 1.00   &  1.87 &  $2.12\pm 0.68$ & 1.00 \\

S/H  &  $6.97_{-0.12}^{+0.15}$ & $7.33_{-0.10}^{+0.11}$ & $6.55 \pm 0.07$ & $6.41 \pm 0.64$ & $6.97_{-0.09}^{+0.11}$ \\

ICF(S) \citepalias{kingsburgh:1994} &  $5.85_{-0.98}^{+1.60}$  &  $9.54_{-1.10}^{+1.60}$    & $2.40_{-0.16}^{+0.21}$  & $1.04\pm 0.28$  & 7.16$_{-0.93}^{+1.19}$ \\

Cl/H &  --- & --- & $6.86_{-0.27}^{+0.21}$ & --- & --- \\ 

ICF(Cl) \citepalias{liu:2000}    &  ---  & ---    & $4.36_{-1.88}^{+2.40}$  & ---  & --- \\ \hline
 
\textit{log(X/O)} &    &  &  &  &  \\

N/O \citepalias{kingsburgh:1994} & $0.98$:: & $1.56$:: & $-0.86 \pm 0.17$ &  $-0.70 \pm 0.47$ & $1.10$:: \\

Ne/O \citepalias{kingsburgh:1994} &  $-0.73 \pm 0.03$ & $-0.61\pm0.02$ & $-0.58_{-0.02}^{+0.03}$ &  $-0.91 \pm 0.78$ & $-0.58\pm0.02$ \\

Ar/O \citepalias{kingsburgh:1994} &  $-2.13_{-0.08}^{+0.07}$ & $-2.32\pm0.03$ & $-2.14 \pm 0.05$ &  $-2.77 \pm 0.87$ & $-2.38\pm0.03$ \\

S/O \citepalias{kingsburgh:1994} &  $-1.21_{-0.08}^{+0.12}$ & $-0.96_{-0.06}^{+0.08}$ & $-1.83_{-0.08}^{+0.07}$ &  $-1.87 \pm 0.65$ & $-1.22 \pm 0.06$ \\

Cl/O \citepalias{liu:2000} &  --- & --- & $-1.49 _{-0.29} ^{+0.19}$    &   --- & --- \\ \hline

\end{tabular}
\label{tab:total-abundances}
\end{table*}

\subsection{Total abundances relative to H (based on ionic abundances relative to H$^+$)}

The abundances of He/H, O/H, N/H, Ne/H, Ar/H, S/H, and Cl/H were calculated from the ionic abundances presented in the subsections above and, the same as in \S 5, the ICFs by \citet{kingsburgh:1994} and by \citet{liu:2000} were used.  The abundance results for observations from 2004 to 2021 are presented in Table \ref{tab:total-abundances} where the abundances relative to O are also listed, in logarithmic scale.
To determine the He abundance only the He$^+$ was considered because no emission from He$^{+2}$ was detected in any of the spectra later than 2000.

The total abundances obtained for all the epochs are sub-solar ($12 + \rm{log(O/H)}$ is between 8.18 and 8.37) and these values are lower than the values reported by \citet{wesson:2005} ($12 + \rm{log(O/H)} = 8.60$) mainly due to the lower temperature used by these authors. 

Due to the very low abundance of O$^+$, determined by using [\ion{O}{ii}] $\lambda$3727+ lines, the ICF for N/H, based on O/O$^+$ is very large and uncertain, highly increasing the N/H value derived from REOSC-Echelle data. These N/H values are very uncertain and have been marked with :: in the tables. The N/H derived from the B\&Ch  data was  obtained by using the ICF O/O$^+$ derived from the auroral lines [\ion{O}{ii}]$\lambda$7325+. This N/H value behaves accordingly with values for non Type I disc PNe. \citet{wesson:2005} derived a $\rm{log(N/O)} = -0.77$ by using the line \ion{N}{iii}]$\lambda$1751 and the ICF $\rm{N/N^{+2} = O/O^{+2}}$, this value is very similar to the N/O values we found for the B\&Ch data.

\section{Ionization Correction Factors}

\label{section:icfs_exp}

ICFs used for the total abundances calculation are listed next. 

\begin{itemize}

\item $\rm{ \frac{He}{H} = \frac{He^{+}}{H^{+}} }$.

\item $\rm{ \frac{O}{H} = \frac{O^+ + O^{+2}}{H^+}}$.

\item $\rm{ \frac{N}{H} = ICF(N) \times \frac{N^+}{H^+}}$. $\rm{ICF(N) = \frac{O}{O^+}}$ \citep{kingsburgh:1994}.

\item $\rm{ \frac{Ar}{H} = ICF(Ar) \times \frac{Ar^{+2} + Ar^{+3} + Ar^{+4}}{H^{+}} }$. $\rm{ICF(Ar) = \frac{1}{1-N^+/N} }$ if Ar$^{+2}$ and Ar$^{+3}$ are detected \citep{kingsburgh:1994}. If only Ar$^{+2}$ is detected: $\rm{ \frac{Ar}{H} = ICF(Ar) \times \frac{Ar^{+2}}{H^+}}$ $\rm{ICF(Ar) = 1.87}$ \citep{kingsburgh:1994}.

\item $\rm{ \frac{Ne}{H} =  ICF(Ne) \times \frac{Ne^{+2}}{H^{+}} }$. $\rm{ICF(Ne) = \frac{O}{O^{+2}} }$ \citep{kingsburgh:1994}.

\item $\rm{ \frac{S}{H} = ICF(S) \times \frac{S^{+} + S^{+2}}{H^{+}} }$. $\rm{ICF(S) = \left[ 1 - \left( 1 - \frac{O^+}{O} \right)^3 \right]^{-1/3}}$. \citep{kingsburgh:1994}.

\item $\rm{ \frac{Cl}{H} = ICF(Cl) \times \frac{Cl^{+2}}{H^+}}$. $\rm{ICF(Cl) = \frac{S}{S^{+2}}}$ \citep{liu:2000}.

\end{itemize}

\section{Atomic data} \label{sec:atomic_data}

The atomic data used in \textsc{PyNeb} calculations are listed in Table \ref{tab:atomic-parameters}.

\begin{table}
\centering
	\caption{\bf Atomic parameters used in \textsc{pyneb} calculations}
 
\begin{tabular}{lcc} \hline
Ion & Transition probabilities & Collisional strenghts \\
\hline
N$^+$ & \citet{froese:2004}  & \citet{tayal:2011}\\
O$^+$ & \citet{froese:2004} & \citet{kisielius:2009}\\
O$^{+2}$ & \citet{froese:2004} & \citet{storey:2014} \\
         & \citet{storey:2000} &  \\
Ne$^{+2}$ & \citet{galavis:1997} & \citet{mclaughlin:2000} \\
S$^+$ & \citet{podobedova:2009} & \citet{tayal:2010} \\
S$^{+2}$ & \citet{podobedova:2009} & \citet{tayal:1999} \\
Cl$^{+2}$ & \citet{mendoza:1983} & \citet{butler:1989} \\
Ar$^{+2}$ & \citet{mendoza:1983} & \citet{galavis:1995} \\
         & \citet{kaufman:1986} &           \\
Ar$^{+3}$ & \citet{mendoza:1982}   & \citet{ramsbottom:1997} \\
          & \citet{kaufman:1986}   & 	\\
Fe$^{+}$  &  \citet{bautista:2015} & \citet{bautista:2015}   \\
Fe$^{+2}$ & \citet{quinet:1996}    & \citet{zhang:1996}  \\
          & \citet{johansson:2000} &           \\
\hline

Ion & \multicolumn{2}{c}{Effective recombination coefficients} \\
\hline
H$^+$ & \multicolumn{2}{c}{\citet{storey:1995}} \\
He$^+$ & \multicolumn{2}{c}{\citet{porter:2012,porter:2013}} \\
N$^{+2}$ & \multicolumn{2}{c}{\citet{fang:2011}}  \\
O$^{+2}$ & \multicolumn{2}{c}{\citet{storey:2017} }  \\
C$^{+2}$ & \multicolumn{2}{c}{\citet{pequignot:1991}}  \\
\hline
\end{tabular}

\label{tab:atomic-parameters}
\end{table}

%%%%%%%%%%%%%%%%%%%%%%%%%%%%%%%%%%%%%%%%%%%%%%%%%%

% Don't change these lines
\bsp	% typesetting comment
\label{lastpage}
\end{document}